\documentclass[12pt]{iopart}

\usepackage[backend=bibtex, maxnames=2,date=year, url=false, doi=false, citestyle=numeric, eprint=true, isbn=false ]{biblatex}

\usepackage{xcolor}
\usepackage{algorithm}
\usepackage{float}
\usepackage[noend]{algpseudocode}
\usepackage{graphicx}
\graphicspath{{images/}} 
\usepackage{caption}
\DeclareCaptionLabelFormat{andtable}{#1~#2  \&  \tablename~\thetable}
\usepackage[labelformat=simple]{subcaption}

\usepackage{amsfonts}
\providecolor{added}{rgb}{0,0,1}
\providecolor{deleted}{rgb}{1,0,0}

\usepackage{xspace}

\makeatletter
\DeclareRobustCommand\onedot{\futurelet\@let@token\@onedot}
\def\@onedot{\ifx\@let@token.\else.\null\fi\xspace}

\def\eg{\emph{e.g}\onedot}

\def\etal{\emph{et al}\onedot}
\makeatother

\newcommand{\changes}[1]{#1}
\begin{document}
	\title[]{Regularising Inverse Problems with Generative Machine Learning Models  }
	\author{M A G Duff$^1$, N D F Campbell$^2$ and   M J Ehrhardt$^1$  
		}
	\address{$^1$ Department of Mathematical Sciences, University of Bath, 
		Bath, BA2\ 7AY, UK}
	\address{$^2$ Department of Computer Science, University of Bath, 
		Bath, BA2\ 7AY, UK}
	\ead{M.A.G.Duff@bath.ac.uk}
	
	\begin{abstract}
		
	Deep neural network approaches to inverse imaging problems have produced impressive results in the last few years. In this paper, we consider the use of generative models   in a variational regularisation approach to inverse problems. The considered regularisers penalise images that are far from the range of a generative model that has learned to produce images similar to a training dataset. We name this family  \textit{generative regularisers}.  The success of generative regularisers depends on the quality of the generative model and so we propose a set of desired criteria to assess generative models and guide future research. In our numerical experiments, we evaluate three common generative models, autoencoders, variational autoencoders and generative adversarial networks,  against our desired criteria. We also test three different generative regularisers   on the inverse problems of deblurring, deconvolution, and tomography. We show that restricting solutions of the inverse problem to lie exactly in the range of a generative model can give good results but that allowing small deviations from the range of the generator produces more consistent results.

	\end{abstract}

	\maketitle
	
	\section{Introduction\label{intro}}
		
Solving an inverse problem is the process of calculating an unknown quantity, $x \in \mathcal{X}$, from observed, potentially noisy, measurements, $y\in \mathcal{Y}$. In this work $\mathcal{X}$ and $\mathcal{Y}$ are assumed to be real finite-dimensional vector spaces.  The two are related by a forward model, $A:\mathcal{X}\rightarrow \mathcal{Y}$, that, for simplicity, is assumed to be linear, giving the equation
\begin{equation}
y=Ax.
\end{equation}
Inverse problems are nearly always ill-posed: there does not exist a unique solution or small deviations in the data lead to large deviations in the solution.
Addressing this is critical for applications where the solution is used to make decisions.
Throughout this paper, we focus on image reconstruction problems,   where  $x \in \mathcal{X}$ is an image, but there are many other applications.

Generally, ill-posed problems are solved by incorporating some prior information; this is often given in the form of a regulariser in a variational regularisation framework~\cite{Scherzer2009, Ito2015, Benning2018}. Consider the optimisation problem
\begin{eqnarray}
x^*\in \arg\min_x L_y(Ax)+\lambda R(x),\label{objective}
\end{eqnarray}
where $L_y:\mathcal{Y} \to [0,\infty] $ is a similarity measure; the constant $\lambda$ is a regularisation parameter  and $R:\mathcal{X}\to [0,\infty] $ is a regulariser that is small when some desired property of the image is fulfilled. For example,  Tikhonov regularisation encourages the reconstruction to be small in the 2-norm, while Total Variation (TV) regularisation~\cite{Rudin1992} allows large gradients (\eg edges) to occur only sparsely in the reconstruction. 
These hand-built regularisers are better suited to some types of images over others, \eg TV is tailored to piece-wise smooth images.  
A natural question to ask is: given a set of images, which regulariser would work well? \changes{Alternatively, how can we produce regularisers that are tailored to specific data or tasks?}

\changes{ There is a wide body of research into learning regularisers. Approaches include using a regulariser to force reconstructions to be sparse in some learned basis for feasible images~\cite{Aharon2006}.  Others have included  a network trained for denoising~\cite{Venkatakrishnan2013, Meinhardt2017, Romano2017} or removing artefacts \cite{ Li2018b, Obmann2020a,   Gonzalez2019} \changes{ in the regulariser term, favouring images that are unchanged by the network}. \ More recently, `adversarial regularisation'~\cite{Lunz2018a} uses a neural network trained to discriminate between desired images and undesired images that contain artefacts.  For a recent overview on  approaches to using deep learning to solve inverse problems, see for example~\cite{Arridge2019}.  }

 In this paper, we consider the case where the regulariser depends on a learned generative model. The	assumption is that the plausible reconstructions exhibit local regularities, global symmetries or  repeating patterns  and so naturally lie on  some lower	dimensional manifold, a subset of $\mathcal{X}$.  A \textit{generator} $G:\mathcal{Z}\rightarrow \mathcal{X}$ takes points from a \emph{latent space}, $\mathcal{Z}$, where  ${\rm dim} (\mathcal{Z})\ll {\rm dim}  (\mathcal{X})$, parameterising this lower dimensional manifold.  In practice, the generator is taken to be a  parameterised function, $G_\theta$, with parameters $\theta$,  for example a  neural network, trained such that the generated points $G_\theta(z) \in \mathcal{X}$ are similar to some pre-defined training set. In this work, we 	 \changes{investigate} regularisers \cite{Bora2017, Dhar2018, Habring2021, Tripathi2018} that penalise values of $x\in\mathcal{X}$ that are far from the range of the generator, $G$, and call these \textit{generative regularisers}.   A popular example \cite{Bora2017}, revisited in Section \ref{Bora-section},  limits solutions to those that are exactly in the range of the generator,
	\begin{eqnarray}
	x^*=G(z^*), \qquad z^* \in  {\rm arg}\min_z \|AG(z)-y\|^2_2+\lambda \|z\|_2^2\label{Bora}.
	\end{eqnarray}  Generative regularisers combine the benefits of both a variational regularisation and data-driven approach. The variational approach builds on the advancements in model-based inverse problems over the last century, while the data-driven approach will provide more specific information than a hand crafted regulariser.  The method remains flexible as the machine learning element is unsupervised and therefore independent of the forward model and the noise type. In this work, we test different generative regularisers, inspired by the literature, on \changes{de}convolution, compressed sensing and tomography inverse problems.	The success of  generative regularisers will depend on the quality of the generator. We propose a set of criteria that would be beneficial for a generative model destined for use in a inverse problems, and demonstrate possible methods of testing  generative models against this criteria.  

\section{Generative Models\label{Gen Models} }

This section provides background on generators and generative models, focusing in particular on three approaches: Autoencoders,  Variational Autoencoders  and Generative Adversarial Networks.

\subsection{Autoencoder (AE)}

An AE has two parts, an encoder and a decoder. The encoder encodes an image in some latent space  and the decoder takes a point in this latent space and decodes it, outputting an image. The lower dimensional latent space forces the network to learn representations of the input with a reduced dimensionality.  Denote the encoder $E_\psi: \mathcal{X}\rightarrow \mathcal{Z}$ and the  decoder  $G_\theta: \mathcal{Z}\rightarrow \mathcal{X}$\changes{,} neural networks with parameters  $\psi$ and $\theta$. The networks are trained by minimising a reconstruction loss \begin{eqnarray}
\mathbb{E}_{x}\left\Vert x-G_\theta(_\psi(x))\right\Vert ^2_2\label{AEs}.
\end{eqnarray} \changes{The expectation is taken empirically over the training dataset}. Post training,  the decoder can be used as a generator. \changes{With no structure imposed on the latent space, generating from random points}  in the latent space may not lead to outputs similar to the training set. Furthermore, points close in the latent space may not lead to similar generated images.  Nevertheless, this method of training is simple and has recently been used in learned singular valued decomposition and for applications in sparse view CT~\cite{Boink2019, Obmann2020}. 

\subsection{Probabilistic Models}

In order to add greater structure and meaning to the latent space and to discourage unrealistic outputs from the generator we consider a  probabilistic approach.  Let $P^*$ be \changes{the} probability distribution of desired solutions to the inverse problem.  We  consider a prior distribution, $P_\mathcal{Z}$, and push it through the generator, $G$, to give a generated  distribution $P_{G}$ on $\mathcal{X}$.  Sampling from the prior and then applying the generator allows samples to be taken from  $P_{G}$.  The generator, $G$,  is chosen to  minimise \changes{a} distance between $P_{G}$ and $P^*$.

\subsection{Generative Adversarial Network (GAN)}
The choice of Wasserstein distance  \changes{between $P_{G}$ and $P^*$}  \changes{and an application of the Kantorovich-Rubinstein duality} leads to the Wasserstein GAN~\cite{Arjovsky2017b}, a popular generative model. Following the derivation given in  \changes{\cite{Arjovsky2017b, Gulrajani2017}}, the task of minimising the Wasserstein distance becomes 
\begin{equation}
\min_{\theta}\max_{\psi}\  \mathbb{E}_{x \sim P^*}D_{\psi}(x)- \mathbb{E}_{z \sim P_\mathcal{Z}} D_{\psi}(G_{\theta}(z))  \label{WGAN}.
\end{equation}
The \textit{generator} $G_\theta: \mathcal{Z}\rightarrow \mathcal{X}$ is as before, and we have introduced a \textit{discriminator} $D_\psi:\mathcal{X}\rightarrow \mathbb{R}$ which must be 1-Lipschitz, enforced by an additional term added to \eref{WGAN}\cite{Gulrajani2017}.

In the  game theoretic interpretation of the GAN~\cite{Goodfellow2014} a generative model competes with a discriminative model.   The discriminator aims to accurately identify real images, maximising $\mathbb{E}_{x \sim P^*} D_{\psi}(x)$, from generated  images, minimising $\mathbb{E}_{z \sim P_\mathcal{Z}} D_{\psi}(G_{\theta}(z))$. The generator tries to force the discriminator to label generated images as real, maximising $\mathbb{E}_{z \sim P_\mathcal{Z}} D_{\psi}(G_{\theta}(z))$.

For the numerical results in this paper  we choose to  use a Wasserstein GAN \eref{WGAN} as it is often more robust to a range of network designs and there is  less evidence of mode collapse, when the generator learns just part of the target distribution, compared to the `vanilla' GAN~\cite{Arjovsky2017b}.

\subsection{Variational Autoencoder (VAE)}
For another choice of distance \changes{between $P_{G}$ and $P^*$,}  take the Kullback--Leibler (KL) divergence ($d_{\rm KL}(P^*\|P_G)= \int_\mathcal{X} \log \frac{dP^*}{dP_G}dP^* $) which leads to  the VAE~\cite{Kingma2014}. Following the derivation in \cite{Kingma2014, Kingma2019}, the \changes{VAE loss function} can be written as 
\begin{eqnarray}
 \mathbb{E}_{x\sim P^*}\left( \mathbb{E}_{z \sim \mathcal{N}_{x, \psi} } \left[ \frac{\| x- G_{\theta}(z) \|_2^2}{2\rho^2}\right]  +d_{KL}\left(\mathcal{N}_{x, \psi}\| P_\mathcal{Z} \right) \right)\label{VAEs}
\end{eqnarray}
where $\mathcal{N}_{x, \psi}:=\mathcal{N}( \mu_{\psi}({x}), {\rm diag}(\sigma^2_{\psi}({x})))$ and $\mu_{\psi},\sigma^2_{\psi}: \mathcal{X}\rightarrow\mathcal{Z}$ are the \textit{encoder mean} and \textit{encoder variance}, neural networks with parameters $\psi$.  The constant $\rho$ is a hyperparameter chosen to reflect the `noise level' in the training dataset.

We can interpret the two terms in \eref{VAEs} as a  data fit and a regulariser term, respectively. In the first term, the   reconstruction and the original image is compared. Encoding to a distribution, $\mathcal{N}_{x, \psi}$, enforces that points close to each other in the latent space should produce similar images.  In the second term, the KL divergence encourages the encoded distributions to be close to the prior, $P_\mathcal{Z}$. The prior is usually taken to be the standard normal distribution. The balance between the two terms is determined by the noise level $\rho$.
 \changes{
\subsection{Other Generative Models }

Generative modelling is a fast growing field and there are other examples of generative models.  Autoregressive models~\cite{Dave2018} generate individual pixels  based on a probability distribution conditioned on previously generated pixels. Normalising flows and, more generally, invertible neural networks~\cite{Jacobsen2018}, map between a latent space and an image space of the same dimension and are designed to be bijective,  with a tractable Jacobian. They can provide a generated distribution with tractable density.  A recent invertible neural network example is a GLOW network~\cite{Kingma2018} which has been used in regularisation of the form $R_G(x)=\|G^{-1}(x)\|_2^2$~\cite{Oberlin}.  Score based generative models, learn to approximate $\nabla p^*$, where $p^*$ is a probability density over the desired image distribution, $P^*$, and can then be used  to sample from $P^*$ using Langevin dynamics \cite{Song2020a}. They have also been used recently as priors for inverse problems, allowing the approximate posterior to be sampled using Monte Carlo methods \cite{Ramzi2020,Jalal2021a}.

 We will consider desired properties of generative models  in more detail in Section~\ref{model_eval}.}
	
	\section{Generative  Regularisers for Inverse Problems \label{Literature Review}}

\changes{

In this section, we bring together current approaches  in the literature  that penalise solutions of an inverse problem that are far from the range of the generator $G$. We consider \changes{variational regularisation \eref{objective} and regularisers }of the form

\begin{equation}
    R_G(x)=\min_{z \in \mathcal{Z}}F(G(z)-x)+ R_\mathcal{Z}(z) \label{gen_regulariser}
\end{equation}
where $F:\mathcal{X}\rightarrow [0,\infty]$ and $R_\mathcal{Z}: \mathcal{Z}\rightarrow [0,\infty]$. We consider  choices for $F$. }

\changes{\subsection{Choices of $F$ in \eref{gen_regulariser}}}

\subsubsection{Restricting solutions to the range of the generator \label{Bora-section}}
\changes{The characteristic function \changes{of} an arbitrary set $\mathcal{C}$ is defined as
		 $$\iota_{\mathcal{C}}(t)=\cases{0&for $t \in \mathcal{C}$\\
			\infty & for $t\notin \mathcal{C}$\\}.	$$
Taking $F(x)= \iota_{\{0\}}(x)$ and $R_\mathcal{Z}(z)=\|z\|_2^2$ in \eref{gen_regulariser} gives  \eref{Bora} and}  describes searching over the latent space for the  encoding that best fits the data. Their choice $R_\mathcal{Z}(z)$ reflects the Gaussian prior  placed on the latent space.  Bora \etal~\cite{Bora2017} first proposed this strategy, applying  it to   compressed sensing problems. There are a number of interesting  applications using  this method, such as  denoising~\cite{Tripathi2018}, \changes{semantic manipulation \cite{Gu2020a}},  seismic waveform inversion~\cite{Mosser2020}, \changes{ light field reconstruction~\cite{Chandramouli2020}}, blind deconvolution~\cite{Asim2018} and phase retrieval~\cite{Hand2018}.  Bora \etal~\cite{Bora2017} assume the existence of an optimisation scheme that can minimise  \eref{Bora} with small error and from this probabilistically bound the \changes{reconstruction error}. However, the non-convexity introduced by the generator  makes any theoretical guarantees on the optimisation extremely difficult.  \changes{Assuming the forward operator is a Gaussian matrix (the generator weights have independent and identically distributed Gaussian entries) and the layers of the generator are sufficiently expansive in size,  there exists theoretical results  on the success of gradient descent for optimising \eref{Bora} \cite{Hand2019,Lei2019a, Daskalakis2020}.}

This formulation \changes{can also be} optimised by projected gradient descent \cite{Shah2018, Jagatap2019}:
\begin{eqnarray}
w_{t+1}=x_t-\eta A^T(Ax_t-y)\label{PGD-1} \nonumber\\ 
z_{t+1}=\arg\min_z\|w_{t+1}-{G}(z)\|_2\label{PGD-2}\\
x_{t+1}={G}(z_{t+1})\nonumber\label{PGD-3}.
\end{eqnarray}
 \changes{With analogies to the restricted isometry property in compressed sensing~\cite{Candes2005},   Shah and Hegde \cite{Shah2018}  introduce the   \textit{Set Restricted Eigenvalue Condition (S-REC)}. If the S-REC  holds,  then the operator $A$ preserves the uniqueness of signals in the range of ${G}$. Theoretical work considers the case where  $A$ is a random Gaussian matrix, and shows, under some assumptions, it satisfies the S-REC with high probability.  In addition, if the generator is an untrained network, then the projected gradient descent approach with sufficiently small step size  converges to $x^*$, where $Ax^*=y$ \cite{Shah2018, Jagatap2019,Peng2019}.}

\subsubsection{Relaxing the Constraints}

Returning to Bora \etal~\cite{Bora2017}, the authors note that as they increase the number of compressed sensing  measurements, the quality of the reconstruction levels off rather than continuing to improve. They hypothesise that this  occurs when the ground truth is not in the range of the generator. One could consider relaxing the constraint that the solution is in the range of the generator, \changes{for example setting $F(x)= ||x||_2^2$ allows for small deviations from the range of the generator. One could also encourage the deviations to be sparse, for example by taking $F(x)= ||x||_1$~\cite{Dhar2018, Hegde2019}.} \changes{Some theoretical considerations for this softly constrained approach is given in~\cite{  Gonzalez2019}. This  approach  is similar to the approaches of~\cite{ Li2018b, Obmann2020a}  where they take $G\circ E: \mathcal{X}\rightarrow \mathcal{X}$  an encoder--decoder network  and define $R_G(x)=\|x-G(E(x))\|_2^2$. The idea is that this regulariser approximates  the distance between $x$ and the ideal data manifold. Less explicitly, there are a number of approaches that extend the range of the original generator, through optimisation of intermediate layers of the network \cite{Menon2020, Daras2021, Gunn2022} or tweaking the generative model training in response to observed data~\cite{Narnhofer2019,Hussein2020}. }

\subsection{Additional regularisation}

Additional regularisation on  $\mathcal{Z}$ is  given by $R_\mathcal{Z}$ in \changes{\eref{gen_regulariser}}. \changes{The most common choice is  $R_\mathcal{Z}(z)=\|z\|_2^2$~\cite{Bora2017,Asim2018} but there are other possibilities, for example} $R_\mathcal{Z}(z)=\iota_{[-1,1]^{d}}(z)$~\cite{Tripathi2018}, where $d={\rm dim} \mathcal{Z}$. Often, the regularisation matches the prior on the latent space used in generator training. Menon \etal~\cite{Menon2020} discuss that  $R_\mathcal{Z}(z)=\|z\|_2^2$   forces latent vectors towards the origin. However, most of the mass of the $d$-dimensional standard normal prior on their latent space is located near the surface of a sphere of radius $d$. Instead,  they	use a uniform prior on $d\mathcal{S}^{{d}-1}$. This idea  has also been explored for interpolations in the latent space~\cite{White2016}. In addition,  the prior on the latent space may  not be a good model for the  post-training subset of $z$ that maps to feasible images. For a  VAE there may be areas of the latent space that the generator has not seen in  training and for  a GAN, there could be mode collapse.  A few recent papers consider how to find the post training latent space distribution~\cite{Dai2019, Bauer2018}.

Other regularisation choices could be based on features of the image, $x={G}(z)$. For example  VanVeen \etal~\cite{VanVeen2018} use  $R_\mathcal{Z}(z)={\rm TV}({G}(z))$. For a GAN generator, it is possible to take the regularisation term to be the same as the generator loss  $R_\mathcal{Z}(z)=\log(1-D( G(z)))$.  This regulariser utilises the discriminator, ${D}$, which has been trained to differentiate generated from real data. Examples include inpainting~\cite{Yeh2017,Lahiri2019} and reconstruction from an unknown forward  model~\cite{Anirudh2018}.

\subsection{Other Approaches\label{exclusions }}


There are a number of  ideas that are linked to earlier discussions in this Section but we will not cover in detail. \changes{A major benefit of \eref{objective} is the flexibility to changes in the forward model. We have threfore ignored conditional generative models~\cite{Adler2018,Park2019a, Yang2018a, lv2021gan, Zhu,Oh2020, Sim2020} and those that train with a specific forward model in mind~\cite{Li2018b, Kabkab2018,Gupta2021}. We also exclude work that uses an untrained neural network, for example Deep Image Priors \cite{VanVeen2018, Ulyanov, Dittmer2019a} or \cite{Habring2021}.}


\section{Generative Model Evaluation\label{model_eval}}

Typically the aim of a generator has been to produce high fidelity images\changes{.} However, the success of \eref{gen_regulariser} relies not just on the ability of the generator to produce a few good images but to be able to produce every possible feasible image. In this section, we discuss desired properties for a generator trained for use in inverse problems and numerically explore  methods to test these properties. 
\subsection{Desired Properties\label{properties}}
	
	To evaluate a generative model, in the context of inverse problems, we consider two overall aims which we will go on to further decompose: 
	\begin{description}
		\item[A] Samples from the generator are similar to those from the target distribution. 
		\item[B]  Given a forward model and an observation, the image in the range of the generator that best fits the observation can be recovered using descent methods. 
		
	\end{description}

	We split aim A into a set of properties: 
	
	\begin{description}
		\item[A1] The generator should be able to produce every possible image in the target distribution. That is, for all  $x \in \mathcal{X}$ such that $x$ is similar to images in the training dataset, there exists $z \in \mathcal{Z}$ such that $G(z)=x$.  
		\item[A2]  The generator should not be able to produce images far from the target distribution. That is, for all  $x \in \mathcal{X}$ such that $x$ is not  similar to images in the training dataset, then there does not exist $z \in \mathcal{Z}$ such that $G(z)=x$.
		
	\end{description}
	
	A1 includes that the generator should be robust to mode collapse and that the model should not trivially over-fit to the training data\changes{.}   
	
	In the probabilistic case, with a prior over the latent space, Property A becomes:
	
	\begin{description}
		\item[A] That samples from the latent space, when mapped through the generator, will produce  samples that approximate a target distribution. We should have that $d(P^*, P_G)$ is small for some distance measure $d$.
	\end{description}
	We also note that in the probabilistic case, 	A1 and  A2  are not independent. By assigning probability mass to parts of the image space close to the target distribution, it is less likely that  images far from the target distribution can be generated. In the probabilistic case, a third Property is added:
	\begin{description}
		\item[A3] The generator should map high probability vectors in the latent space distribution to high probability images in the target distribution.  
	\end{description}
	It is possible that A1 and  A2 are satisfied but not A3.  Note that these properties may not be possible to achieve for a given dataset. 
	
	We define two properties for Property B, these are
	\begin{description}
		\item[B1] The generator should be  smooth with respect to the latent space, $\mathcal{Z}$. 
		\item[B2] The area of the latent space, $\mathcal{Z}$, that corresponds to  images  \changes{similar to those in the training set} should be known.

	\end{description}	
	
	B1 ensures that gradient--based optimisation methods can be used. Continuity is also desirable: we wish that, in some way, points close in the latent space should produce similar images. B2 considers that we need to have a distribution on or subset of $\mathcal{Z}$ to sample from in order to use the generator to sample images. This distribution may not \changes{necessarily} be equal to any priors on the latent space used during training.  \changes{We recognise that B1 and B2 are perhaps vague, and are potentially not sufficient for Property B. It is an area for future work to consider making these statements precise enough to support theoretical work. }

\subsection{Generative Model Evaluation Methods}

There are a wide range of existing generative model evaluation methods~\cite{Borji2019}, focused mostly on Property A. We assume the availability of some test data drawn from the same distribution as the training data and unseen by the generative model.  The \textit{average log likelihood}~\cite{Goodfellow2014} of  test data under the generated distribution is a natural objective to maximise. There is evidence, however,    that the likelihood is generally unrelated to image quality \changes{and is difficult to approximate in higher dimensions~\cite{Theis2015b}}.  \changes{To calculate a distance between generated and desired distributions, one possibility is the earth movers  distance (EMD)~\cite{Rubner1998}, a discretised version of the Wasserstein distance. One could also encode the generated and unseen data in a lower dimensional space before taking distance calculations, for example by taking the outputs of one layer of any neural network trained for classification~\cite{Heusel2017,Gretton2012}.  A model that overfits to the training data would perform perfectly in these distance measures.  Also, the low dimensional representation used for the evaluation is likely to have the same inherent problems and drawbacks as the embedding learnt by the generative model.}   Similarly, a number of tests train an additional, separate, neural network discriminator to distinguish between  test data and generated \changes{data \cite{Arjovsky2017b,Lopez-Paz2019}}. Failure to classify the two is a success of the generative model. For testing a GAN,  the new discriminator  is unlikely to be able to pick up failures that the original discriminator, used in training, missed.  Finally, Arora \etal~\cite{Arora2017}  estimate the size of the support of the generated distribution.  A low support size would suggest  mode collapse. \changes{Their technique depends on manually finding duplicate generated images which can be time consuming and require expert knowledge. }

Property B is less explored in the literature.  \changes{One approach is to directly attempt to reconstruct test data by finding a latent space vector that when pushed through the generator, matches the data.}  With these found latent vectors, analysing  their locations could check Property B2.  To test the smoothness of the generator with respect to the latent space, Property B1, many previous papers, including the original GAN and VAE papers \cite{Goodfellow2014,Kingma2014},  interpolate through the latent space, checking for smooth transitions in the generated images.

\subsection{Numerical Experiments\label{numerics}}

In this Section, we evaluate  AE, VAE and GAN models  against the desired properties given in Section~\ref{properties}. We consider experiments on  two datasets. Firstly, a custom made \texttt{Shapes} dataset with 60,000 training and 10,000 test $56\times56$ grey-scale images. Each image consists of a black background with a grey circle and rectangle of constant colour. The radius of the circle; height and width of the rectangle; and locations of the two shapes, are sampled uniformly with ranges chosen such that the shapes do not overlap.  This dataset is similar to the one used in~\cite{Park2019a}.  Secondly, the \texttt{MNIST} dataset~\cite{LeCun1998} consists of $28\times28$ grey-scale images of handwritten digits with a training set of 60,000 samples, and a test set of 10,000 samples. For examples of both datasets, see the ground truth images in  Figure  \ref{fig:mnist_shapes_reconstructions}.

\changes{ Architecture details are given in the appendix. We chose to use the same generator network for all three models, for comparison. Architecture choices were guided by~\cite{Kingma2014,Gulrajani2017,Salimans2016}.}   All models have gone through a similar amount of  optimisation of  hyperparameters, including: the noise level $\rho$ in the VAE decoder \eref{VAEs}; the latent dimension;  number of layers; choice of convolution kernel size; drop out probability; leaky ReLU coefficient and  learning rate. In order to select hyperparameters we manually inspected generated images. Models were built and trained using Tensorflow~\cite{tensorflow2015-whitepaper} in Python and made use of  the Balena High Performance Computing  Service at the University of Bath. The models were trained using a single Dell PowerEdge C8220X node, with two Intel E5-2650 v2 CPU, 	64 GB DDR3-1866 MHz Memory and an Nvidia K20X GPU, 6 GB memory. The \texttt{MNIST} and  \texttt{Shapes} VAE models taking approximately 25 and 45 minutes to train, respectively.

\subsubsection{Reconstructing a Test Dataset\label{reconstruction}}

Property A1 asks that the generator is able to produce every image in the target distribution. Gradient descent with backtracking  line search (Algorithm \ref{GD}, in the appendix)  is used to approximate
\begin{eqnarray}
z^*(x) \in \arg\min_z \|G(z)-x\|_2^2,\label{encode}
\end{eqnarray}
for each $x \in \mathcal{X}_{{\rm test}}$, an unseen test dataset. For the AE and VAE, the algorithm is initialised at the (mean) encoding  of the test image, $E_\psi(x)$ and $\mu_\psi(x)$, respectively. For the GAN, \changes{we take 4 different initialisations, drawn from a standard normal distribution, and take the best result.  }  We find empirically, especially for the GAN, that different initialisations lead to different solutions. 

\changes{
 Figure~\ref{fig:violinplots} shows  \changes{ $\|G(z^*(x))-x \|_2 / \|x\|_2$, the normalised root mean squared error (NRMSE) for reconstructions on \texttt{Shapes} and \texttt{MNIST}  for the three different generator models.} We see that,  for both datasets, the AE and VAE have almost identical reconstruction results and the GAN results are comparatively worse.  For the \texttt{Shapes} dataset the difference in results between the three generative models is less stark. In addition, NRMSE values are given for three different latent dimensions to show that the results are not sensitive to small changes in latent dimension. Latent dimensions of 8 and 10 for \texttt{MNIST} and \texttt{Shapes}, respectively,   are used in the rest of this paper.  Figure~\ref{fig:mnist_shapes_reconstructions} also shows reconstruction examples \changes{providing context to the results in  Figure~\ref{fig:violinplots}. } Numerical values on the image use the Peak-Signal-to-Noise-Ratio  (PSNR, see definition 3.5 in~\cite{Bredies2018}).     The non-circular  objects  in the  GAN results for \texttt{Shapes}  could be a failure of the discriminator to detect circles.}
 
\changes{
\begin{figure}
\centering
	\begin{subfigure}{.49\textwidth}
		\centering
		\includegraphics[width=\linewidth]{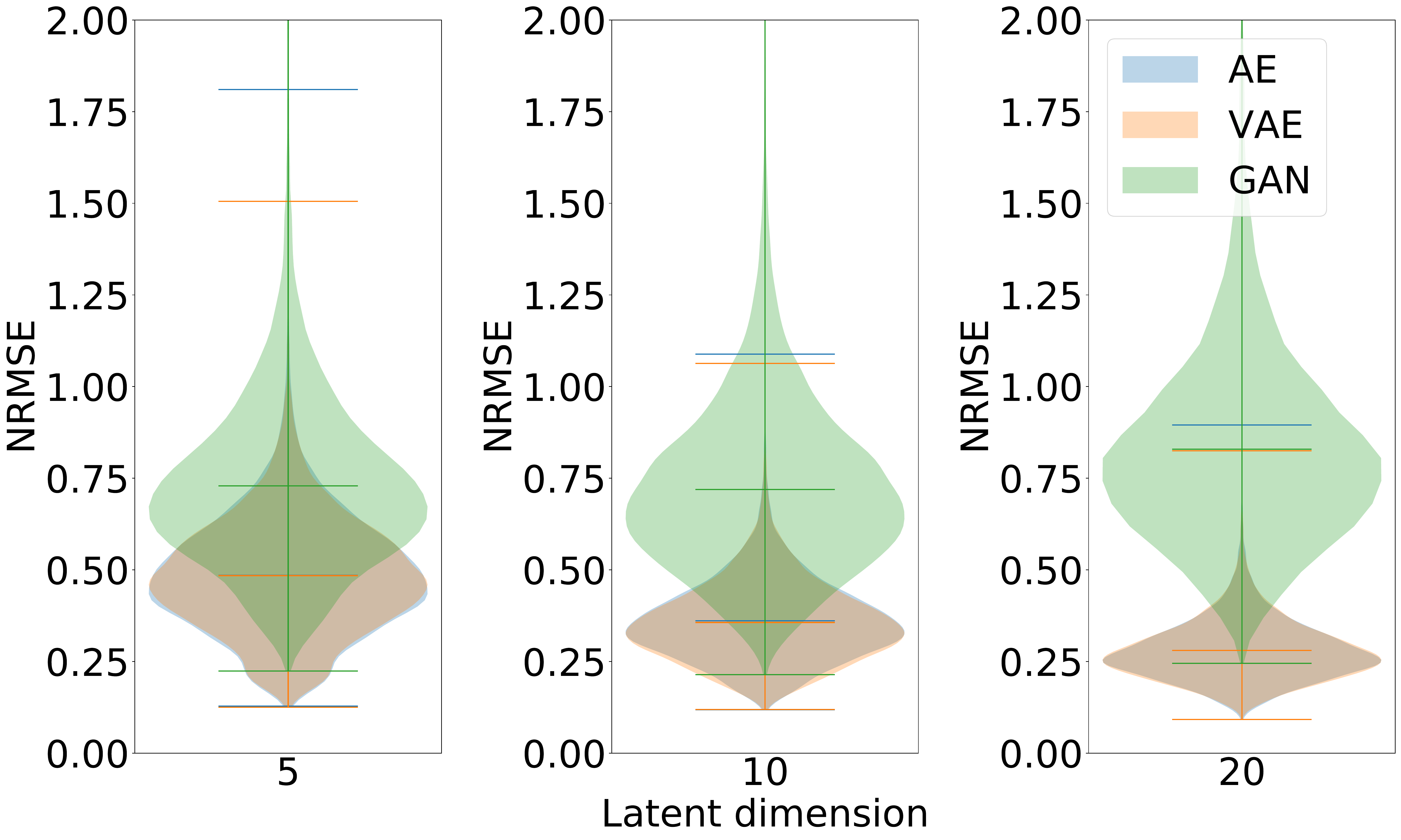}  
		\caption{\texttt{MNIST} dataset}
		\label{fig:violinplots-mnist}
	\end{subfigure}
	\begin{subfigure}{.49\textwidth}
		\centering
		\includegraphics[width=\linewidth]{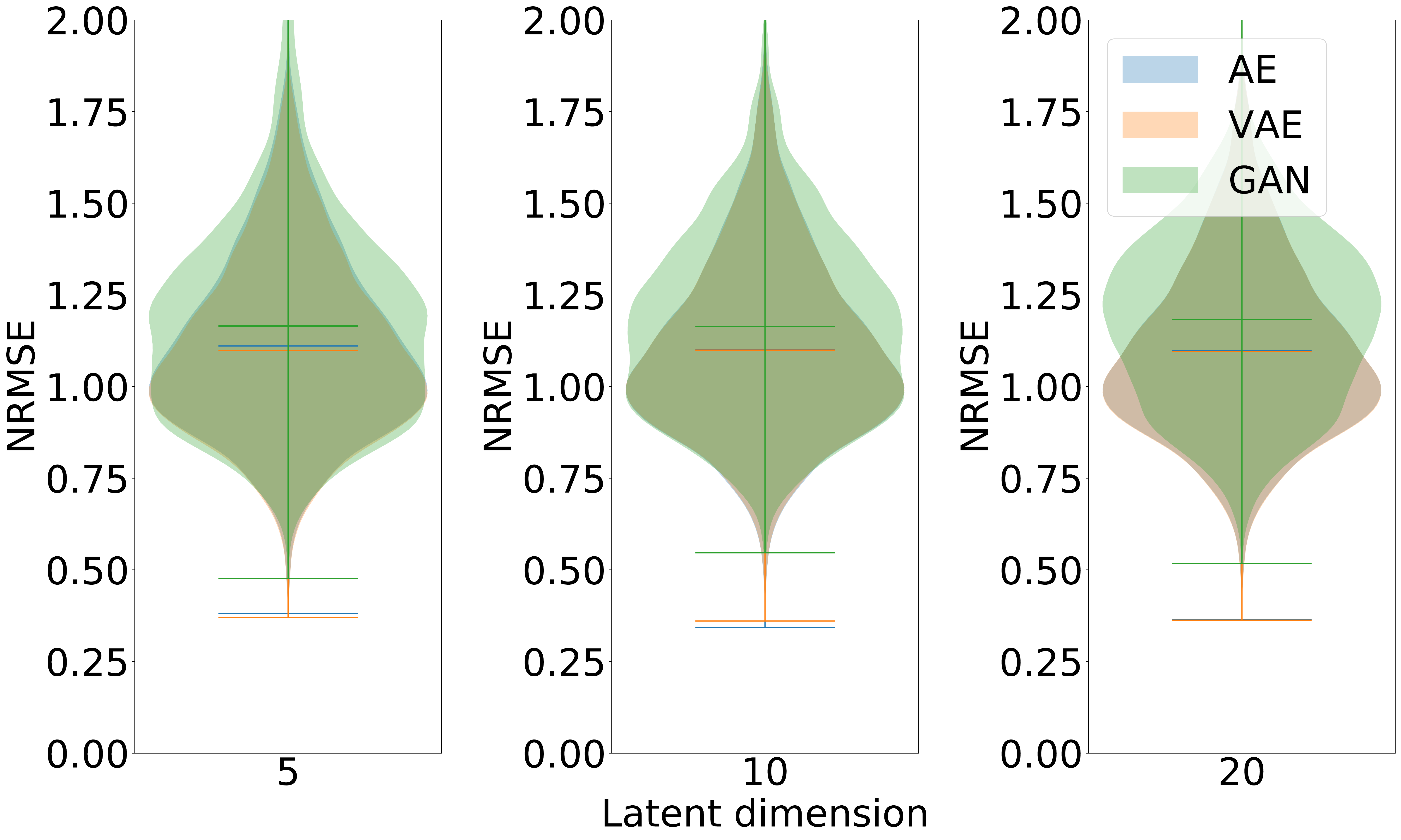}  
		\caption{\texttt{Shapes} dataset}
		\label{fig:violinplots-2dshapes}
	\end{subfigure}
	\caption{\changes{NRMSE between values of $G\left(\arg \min_z\|G(z)-x\|_2\right)$ and  $x$  and plotted as a histogram for all $x\in \mathcal{X}_{\rm test}$. The horizontal lines show the median and range and the shaded area a histogram.  Note the brown colour is the result of the overlapping orange (VAE) and blue (AE).}}
	\label{fig:violinplots}
	
			\centering
		\begin{subfigure}{.315\textwidth}
		\includegraphics[width=\linewidth]{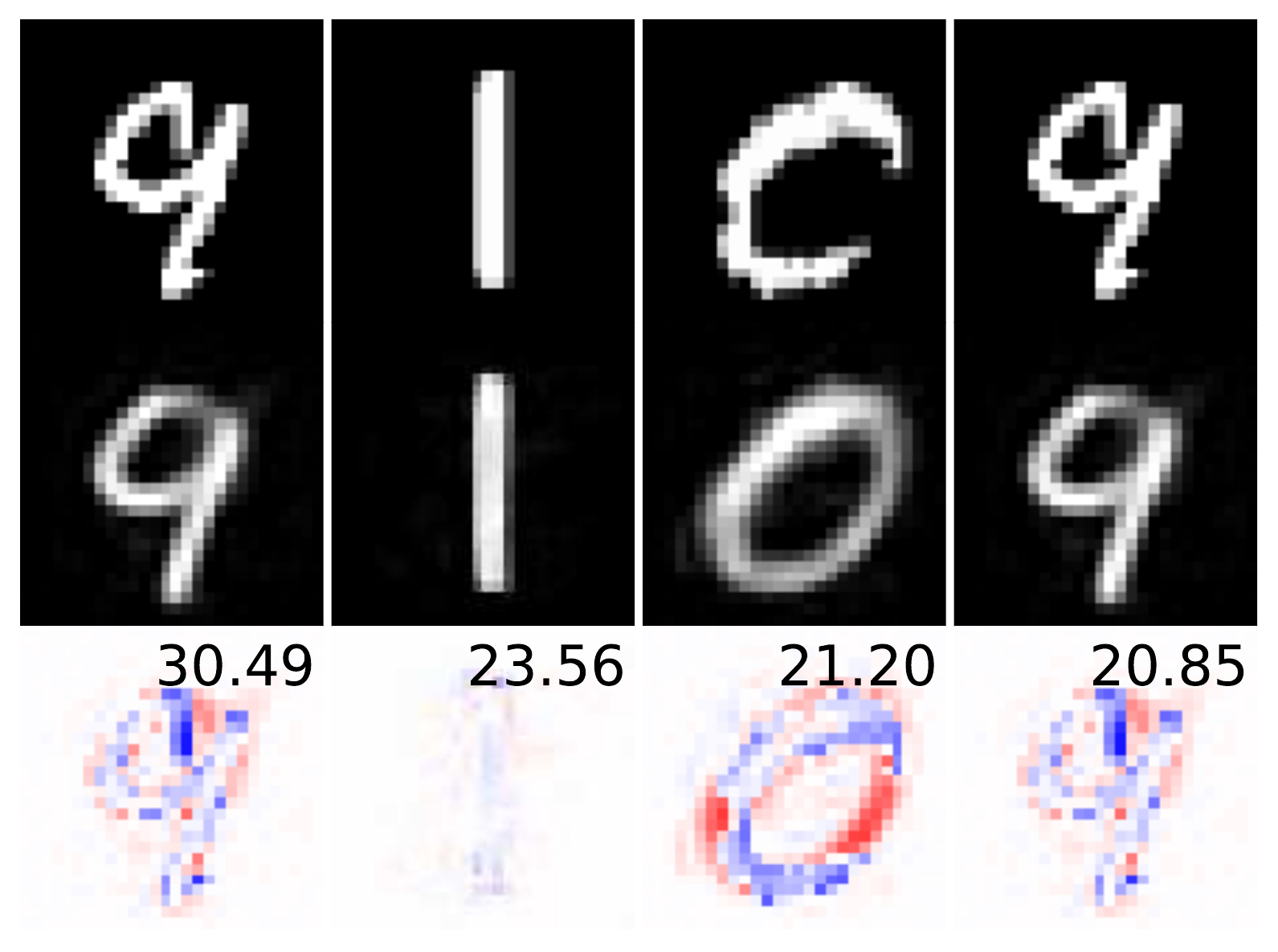}
		\caption{AE -- \texttt{MNIST}}
			\end{subfigure}
			\begin{subfigure}{.315\textwidth}
		\includegraphics[width=\linewidth]{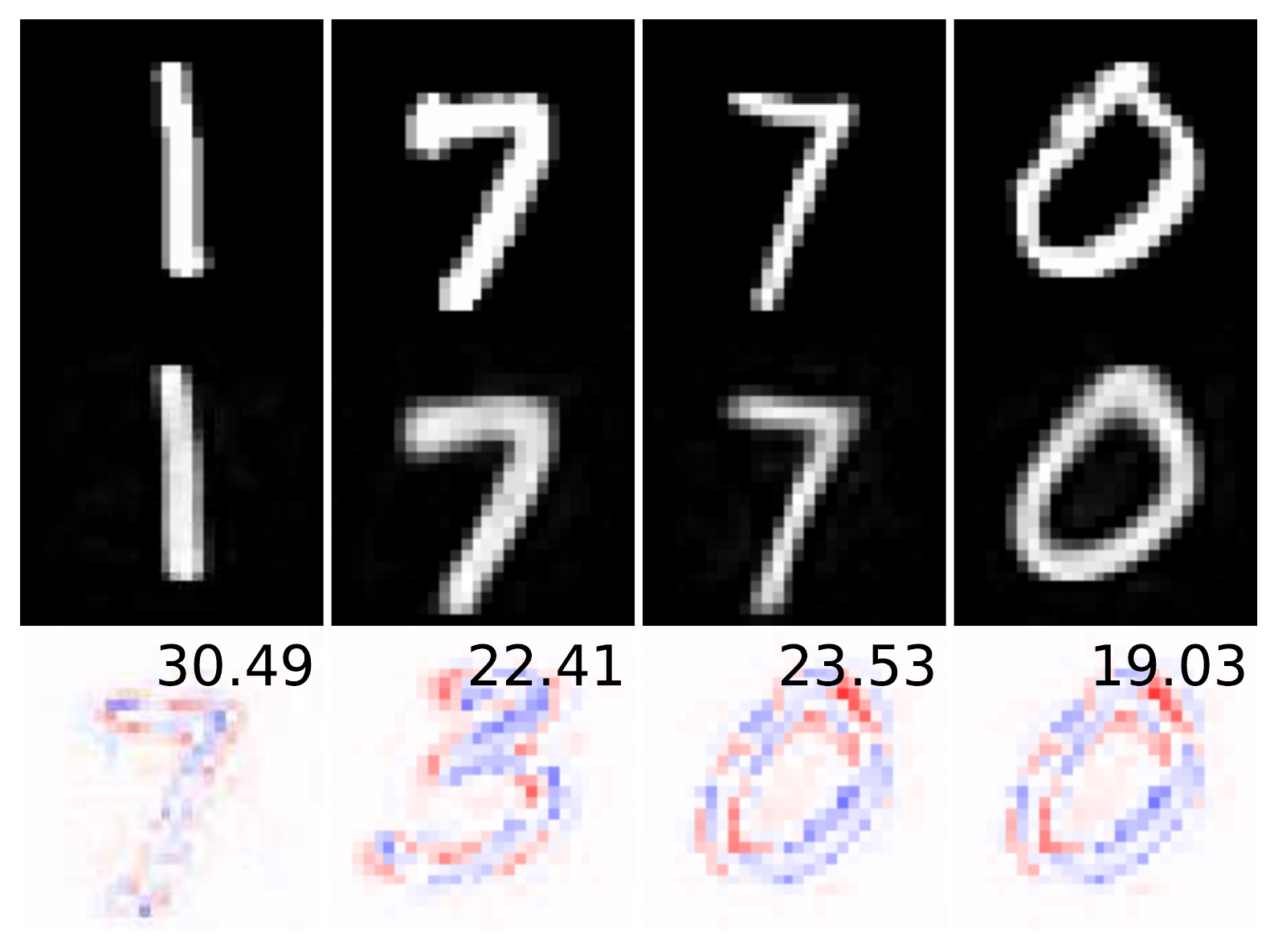}
				\caption{VAE -- \texttt{MNIST}}
			\end{subfigure}
			\begin{subfigure}{.35\textwidth}
		\includegraphics[width=\linewidth]{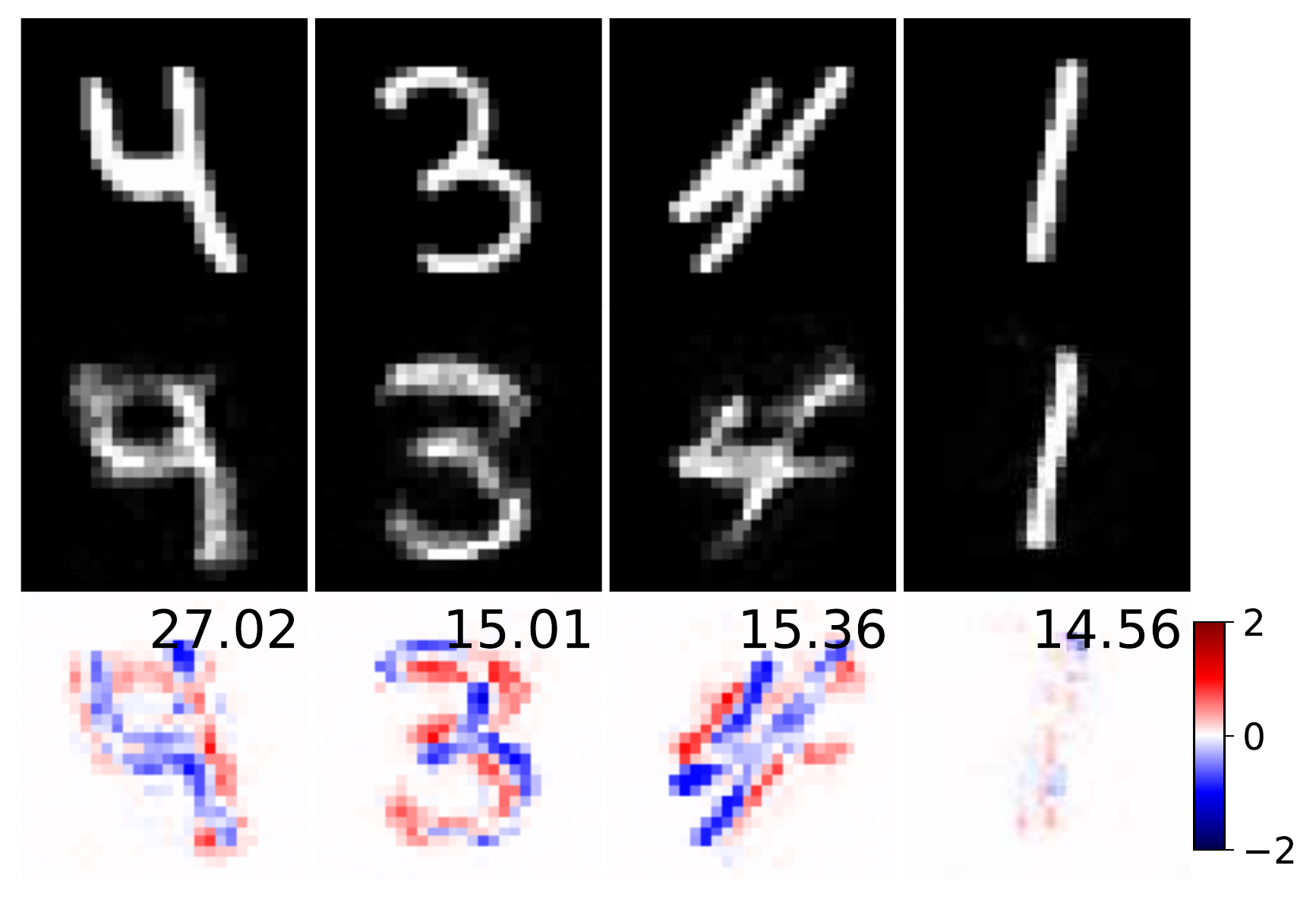}
				\caption{GAN -- \texttt{MNIST}}
			\end{subfigure}
			
		\centering
		\begin{subfigure}{.315\textwidth}
		\includegraphics[width=\linewidth]{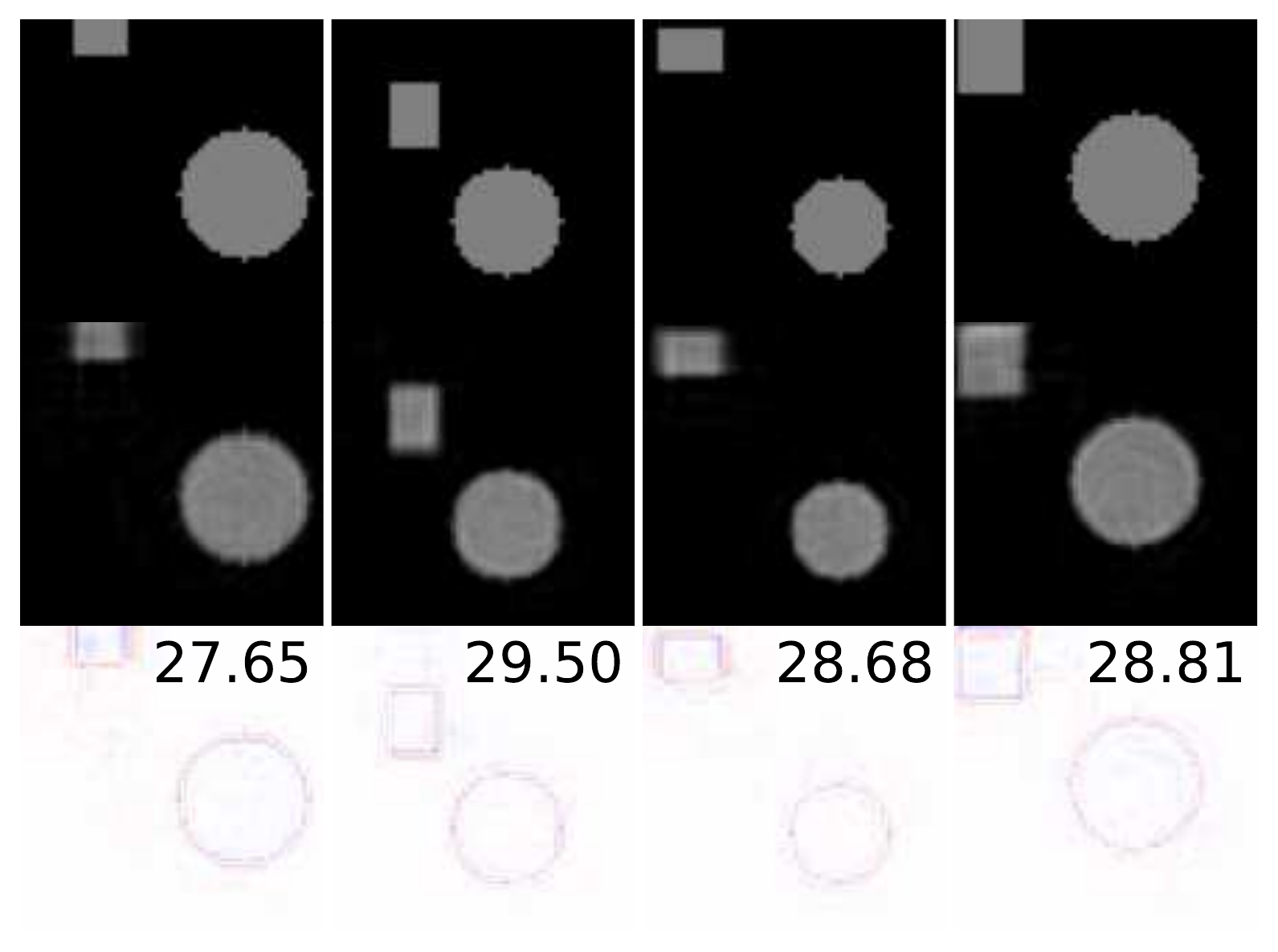}
		\caption{AE -- \texttt{Shapes}}
			\end{subfigure}
			\begin{subfigure}{.315\textwidth}
		\includegraphics[width=\linewidth]{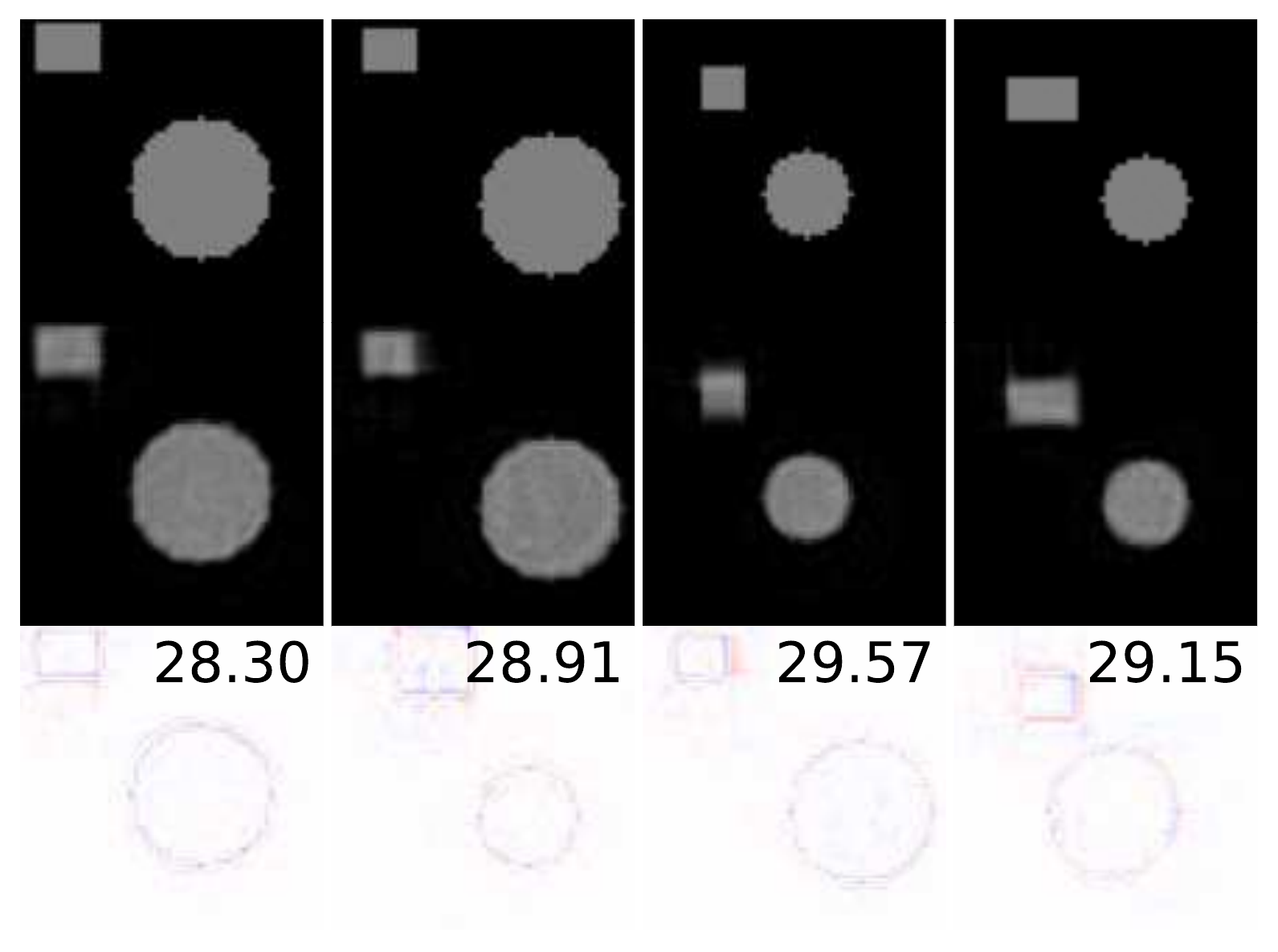}
				\caption{VAE -- \texttt{Shapes}}
			\end{subfigure}
			\begin{subfigure}{.35\textwidth}
		\includegraphics[width=\linewidth]{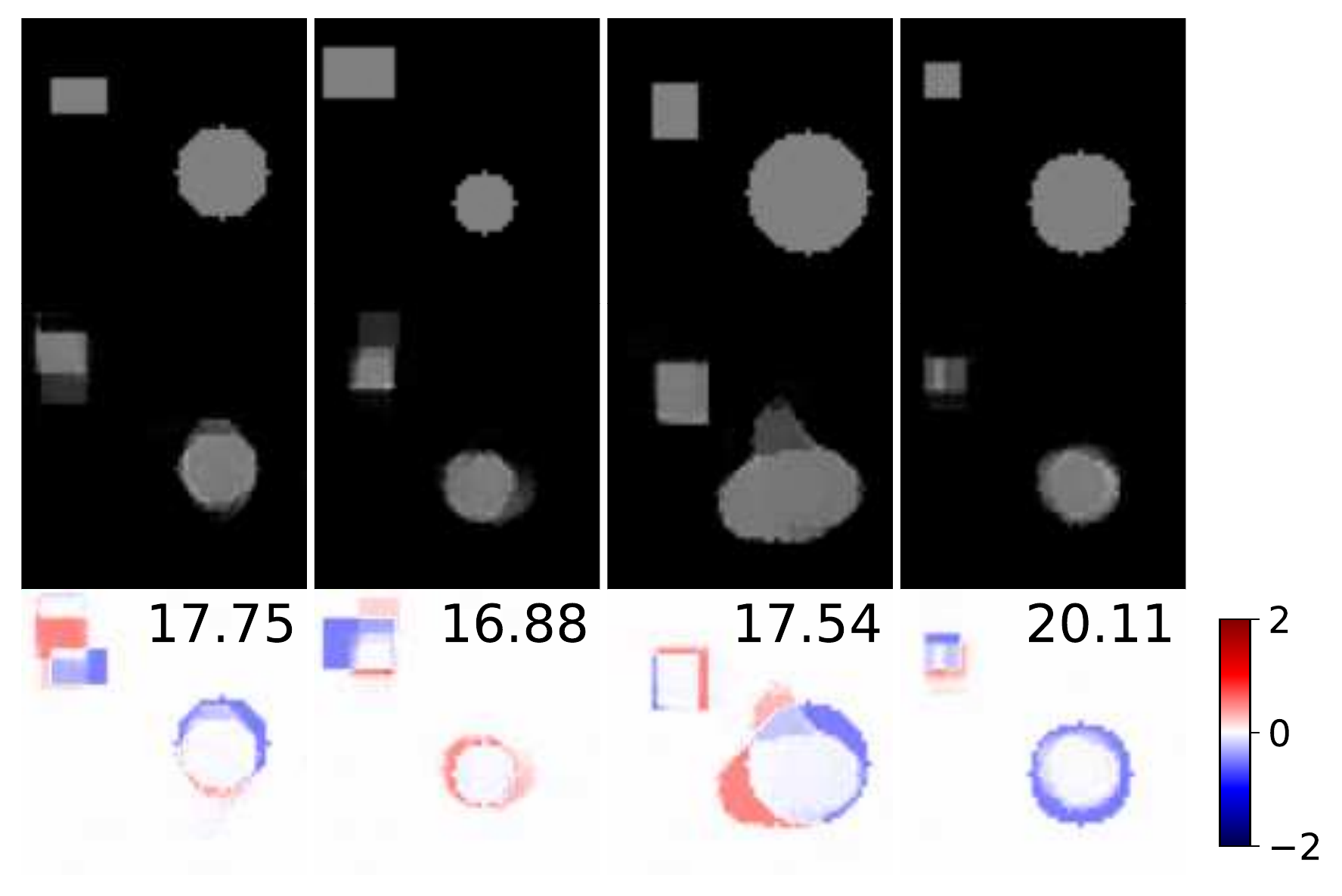}
				\caption{GAN -- \texttt{Shapes}}
			\end{subfigure}
			
	\caption{\changes{Example reconstructions for the \texttt{MNIST} and \texttt{Shapes} dataset with  eight ten-dimensional generative models respectively. In each sub-figure, the top row shows the ground truth, second row the reconstruction and  third row the difference between the two.}}
	\label{fig:mnist_shapes_reconstructions}
	
\end{figure}}

\subsubsection{Distance Between $P_G$ and $P^*$\label{emd}}

\begin{figure}
\centering 
	\begin{subfigure}{.35\textwidth}
		\centering
		\includegraphics[width=\linewidth]{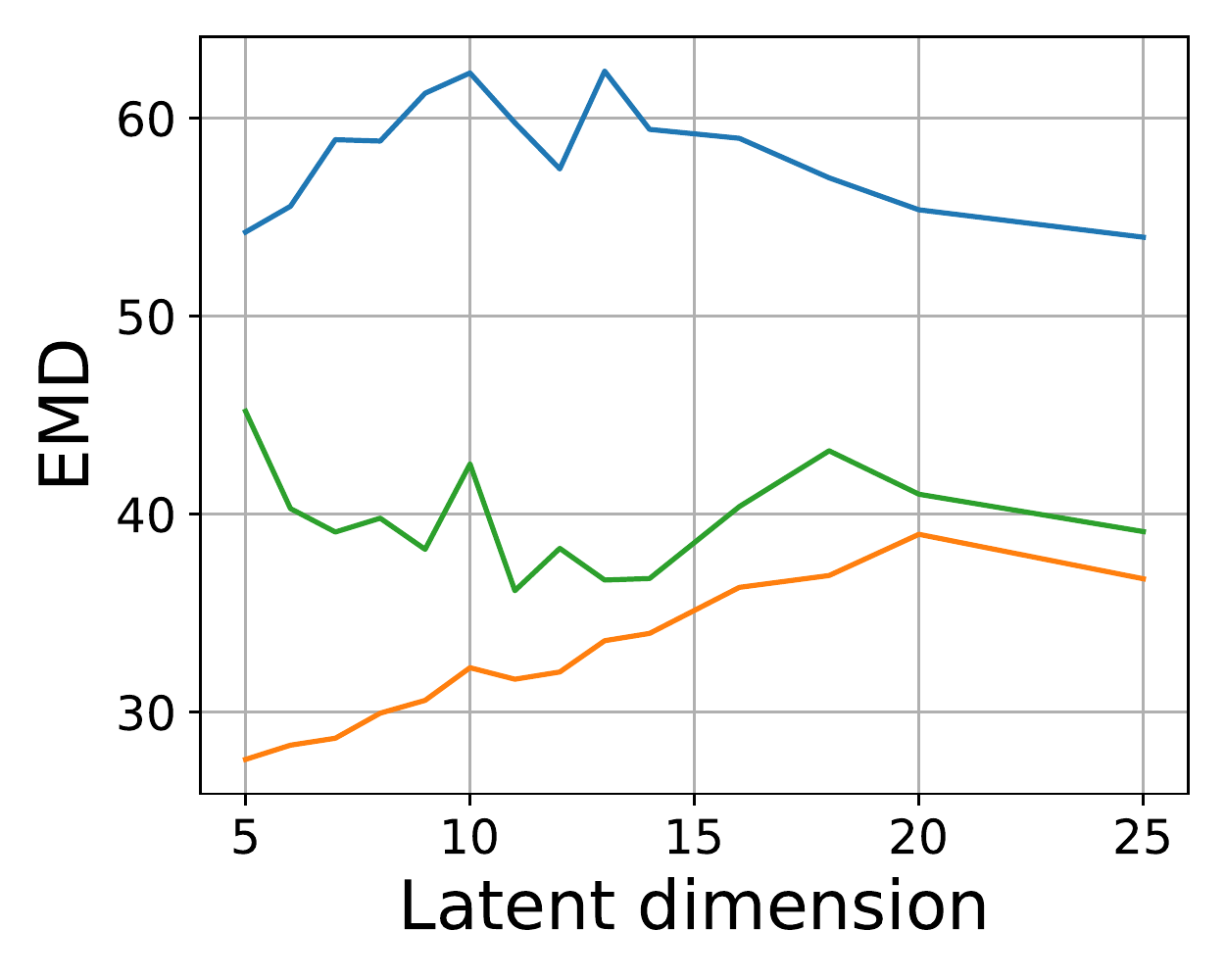}  
		\caption{ \texttt{MNIST}}
		\label{fig:mnistemd}
	\end{subfigure}
	\qquad\qquad
	\begin{subfigure}{.35\textwidth}
		\centering
		\includegraphics[width=\linewidth]{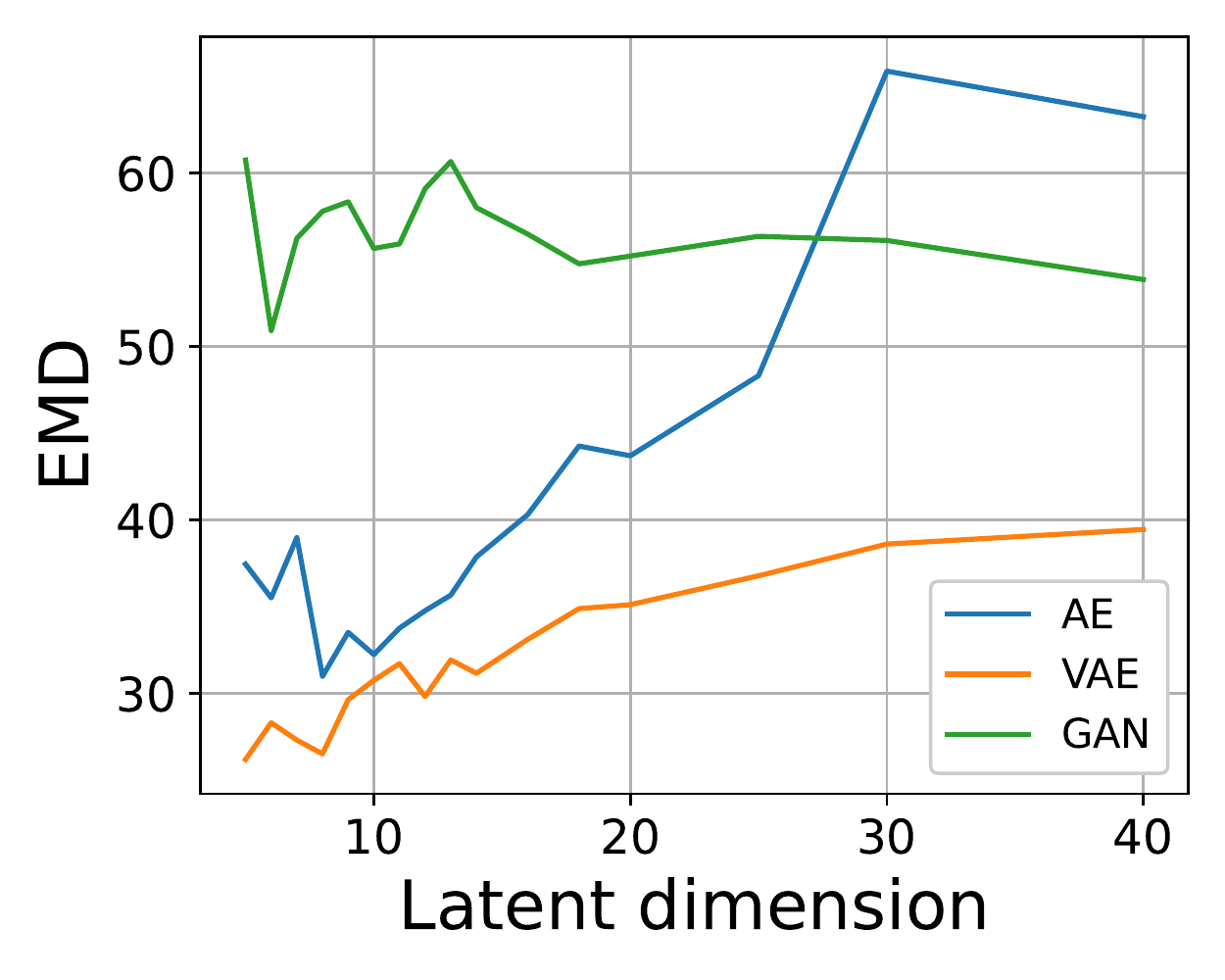}  
		\caption{\texttt{Shapes}}
		\label{fig:2d_shapesemd}
	\end{subfigure}
	\caption{EMD between the test dataset and samples from a trained generator.}
	\label{fig:distributionscompare}
\end{figure}

\changes{To investigate Property A3, the EMD~\cite{Rubner1998} is calculated between empirical observations of the generated and the data distributions $P_G$ and $P^*$. For the sets of test and generated images, $\mathcal{X}_{{\rm test}} =\{x_1,\ldots,x_N\}$ and $\{G(z_1), \ldots, G(z_N) : z_i \sim \mathcal{N}(0,I)\}$, the EMD between their empirical distributions is defined as
\begin{eqnarray}
 \min_f \left\{ \sum_{i,j=1}^{N}f_{i,j}\|x_i-G(z_j)\|^2_2 : 0\leq f_{i,j}\leq 1,\sum_{i=1}^N f_{i,j}=1, \sum_{j=1}^N f_{i,j}=1\right\}.
\end{eqnarray}}
The EMD is calculated using the Python Optimal Transport Library~\cite{Flamary} with $N=10,000$, the full test set.  The results are given in Figure~\ref{fig:distributionscompare}. In both the \texttt{MNIST} and  \texttt{Shapes} examples, the VAE \changes{has a lower EMD across the latent dimensions.} The AE is added to this plot for comparison purposes but, as there is no prior on the latent space,  $z_i \sim \mathcal{N}(0,I)$ may not be a suitable choice to sample from.

\subsubsection{	Visualisations of the Latent Space\label{visualisation_latent}}

\begin{figure}
	\begin{subfigure}{.49\textwidth}
		\centering
		\includegraphics[width=\linewidth]{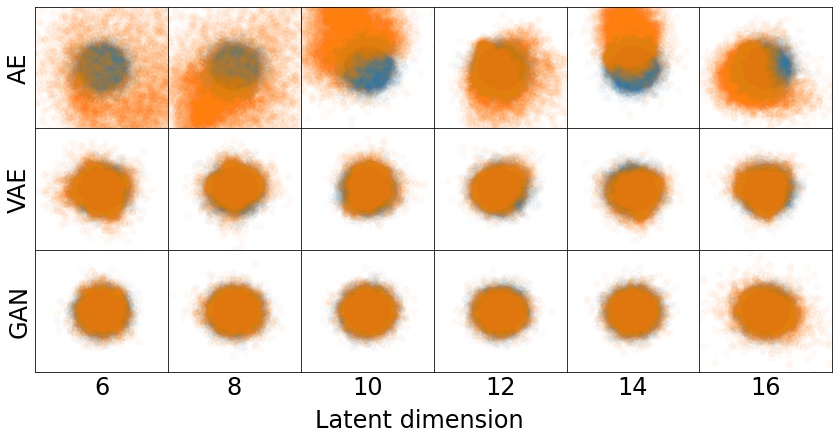}  
		\caption{\texttt{MNIST}}
		\label{fig:mnistencodingprojectionplots}
	\end{subfigure}
	\begin{subfigure}{.49\textwidth}
		\centering
		\includegraphics[width=\linewidth]{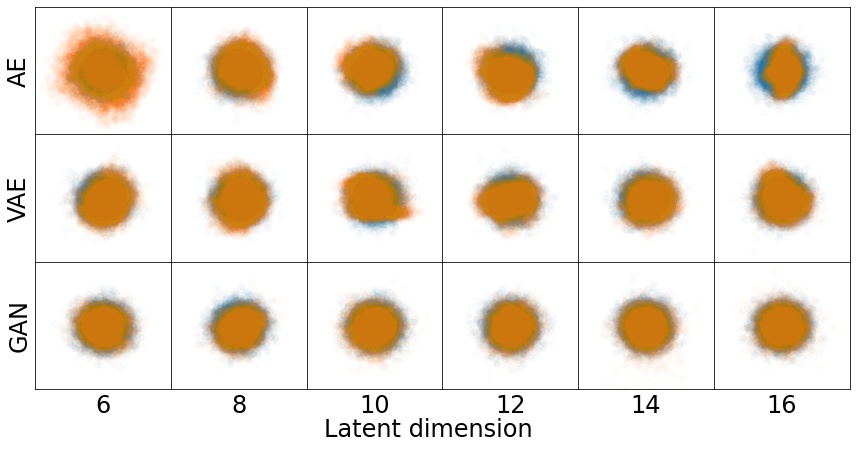}  
		\caption{\texttt{Shapes}}
		\label{fig:2dshapesencodingprojectionplots}
	\end{subfigure}
	\caption{Comparisons of the latent space encodings of a test dataset with a standard normal distribution by projecting the vectors into 2 dimensions. Encodings of the test dataset are  in orange and  the standard normal vectors in blue.}
	\label{fig:encodingprojectionplots}
\end{figure}

Property B2 requires that the area of the latent space that maps to feasible images is known.  There is no prior on the latent space enforced for  AEs and a $\mathcal{N}(0,I)$ prior is imposed for VAEs and GANs. In Figure~\ref{fig:encodingprojectionplots}, gradient descent with backtracking (Algorithm \ref{GD} in the appendix) is used to approximate \eref{encode}, finding a latent vector $z^*(x)$ for each $x \in \mathcal{X}_{{\rm test}}$. For comparison, the values  $z^*(x)$ for the test set and 10,000 vectors drawn from a standard normal distribution are  randomly projected into 2 dimensions. The encodings in the latent space match the prior $\mathcal{N}(0,I)$  for VAEs and GANs.  For AEs, there are examples in lower latent dimensions, where the area covered by the encodings does not match a standard normal distribution. 

	\subsubsection{Generating Far from the Latent Distribution}

\begin{figure}[h]
	\centering
	\begin{subfigure}{.49\textwidth}
		\centering
		\includegraphics[width=\linewidth]{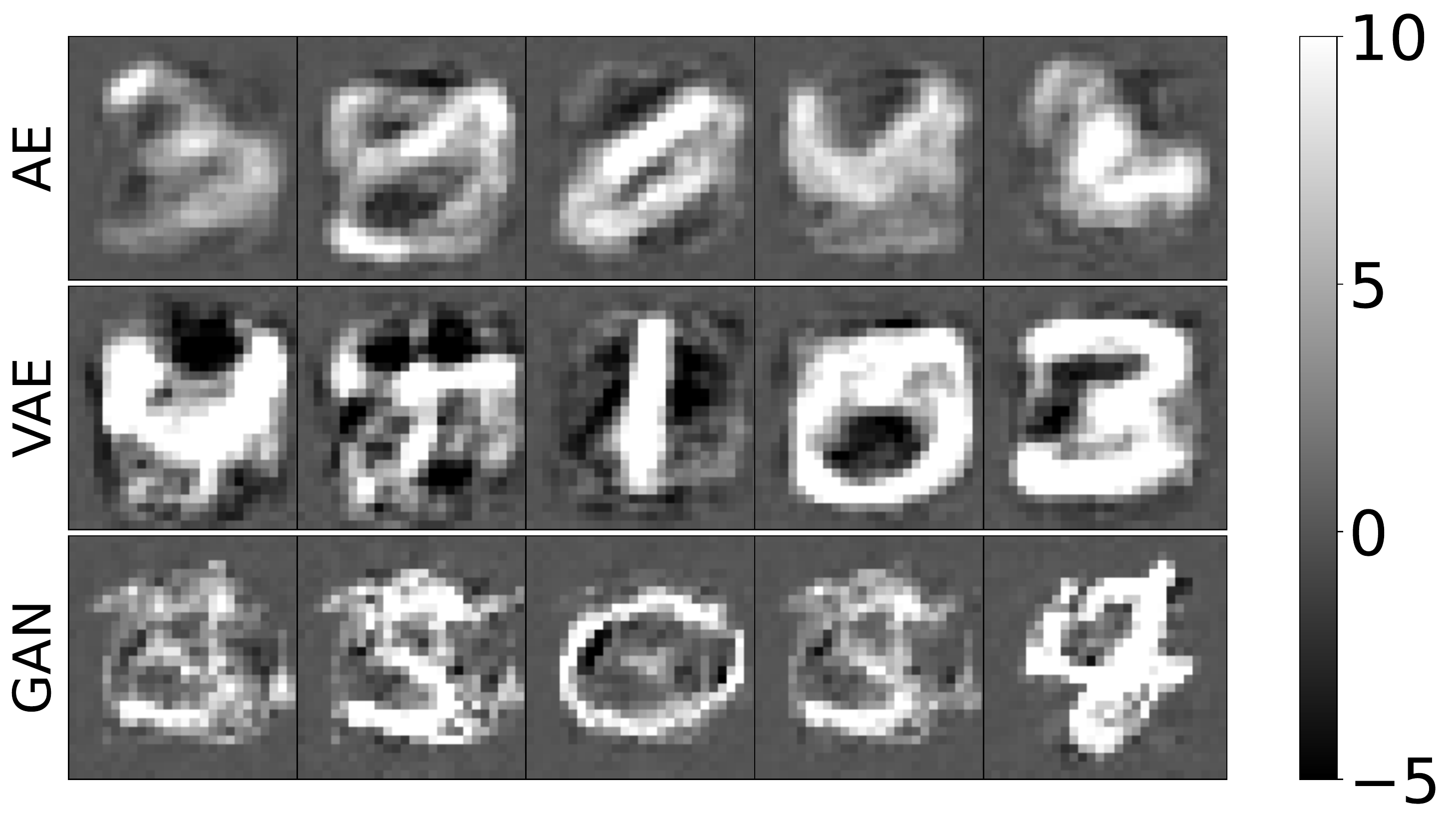}
		\caption{\texttt{MNIST}}
		\label{fig:mnist10dimfarfromdistribution}
	\end{subfigure}
	\hfill
	\begin{subfigure}{.49\textwidth}
		\centering
		\includegraphics[width=\linewidth]{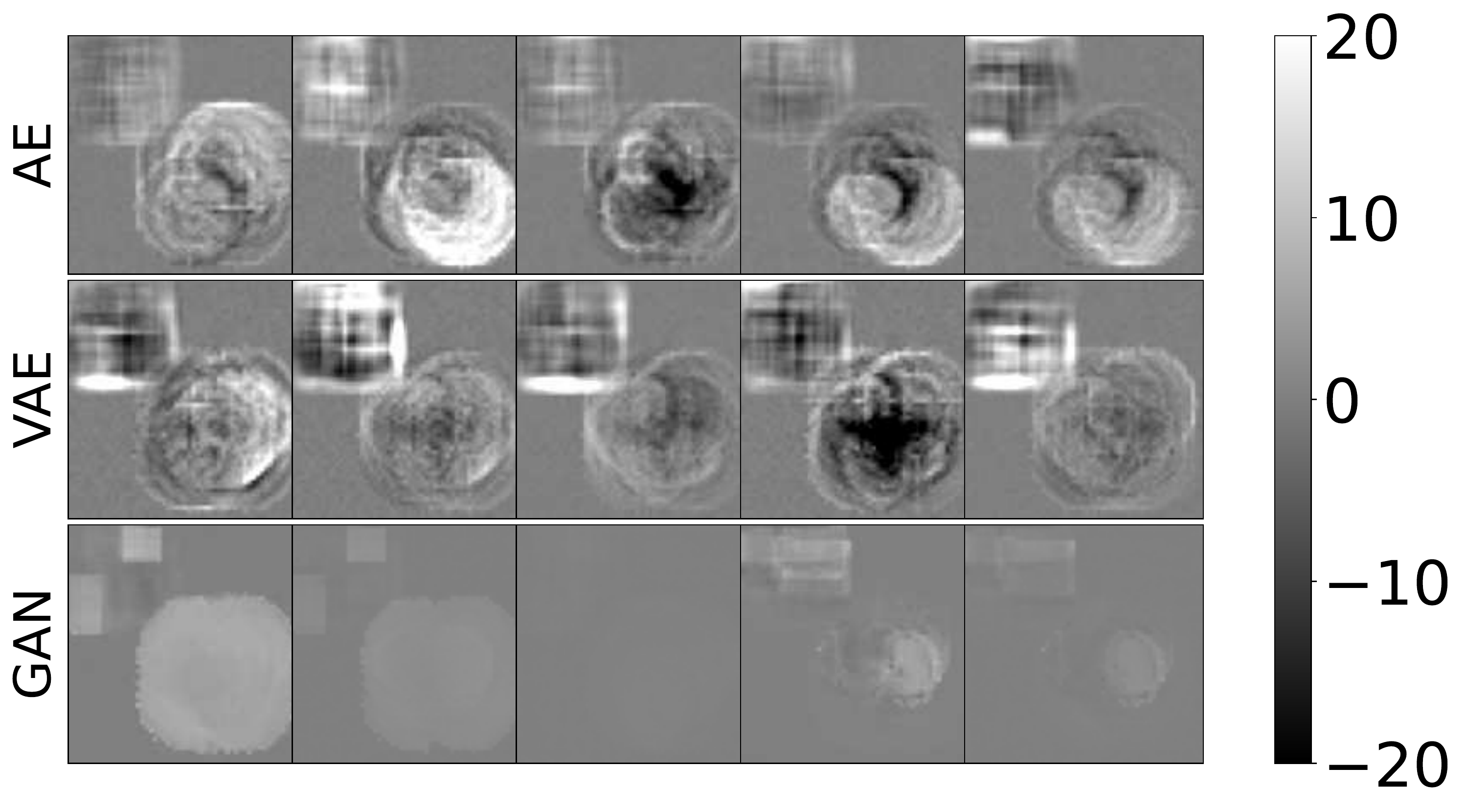}
		\caption{\texttt{Shapes}}
		\label{fig:2dshapes10dimfarfromdistribution}
	\end{subfigure}
	\caption{Images generated far from the high-probability region of the prior distribution.}
	\label{fig:farfromdistribution}
\end{figure}

A known latent space gives known areas to sample from to produce new images.   Figure~\ref{fig:farfromdistribution} shows image examples generated far from \changes{a standard normal distribution}. \changes{The images are not recognisable as similar to the training datasets.  This emphasises the importance of Property B2, that the area of the latent space that corresponds to  images  \changes{similar to those in the training set} should be known. }

	\subsubsection{Interpolations in the Latent Space}
\begin{figure}
	\centering
	\begin{subfigure}{.32\textwidth}
		\centering
		\includegraphics[width=\linewidth]{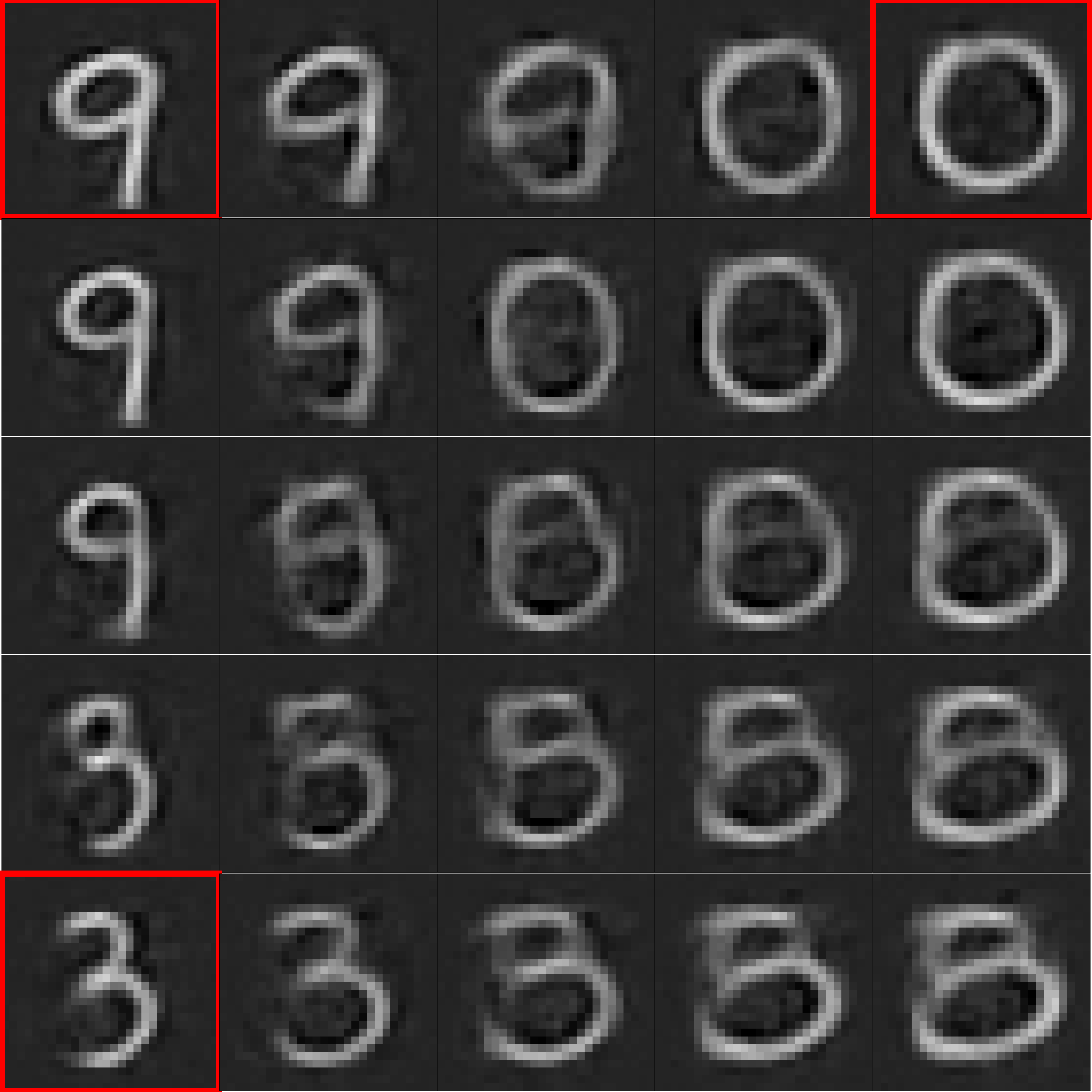}  
		\caption{\texttt{MNIST} - AE}
		\label{fig:aemnistinterpolations8}
	\end{subfigure}
	\begin{subfigure}{.32\textwidth}
		\centering
		\includegraphics[width=\linewidth]{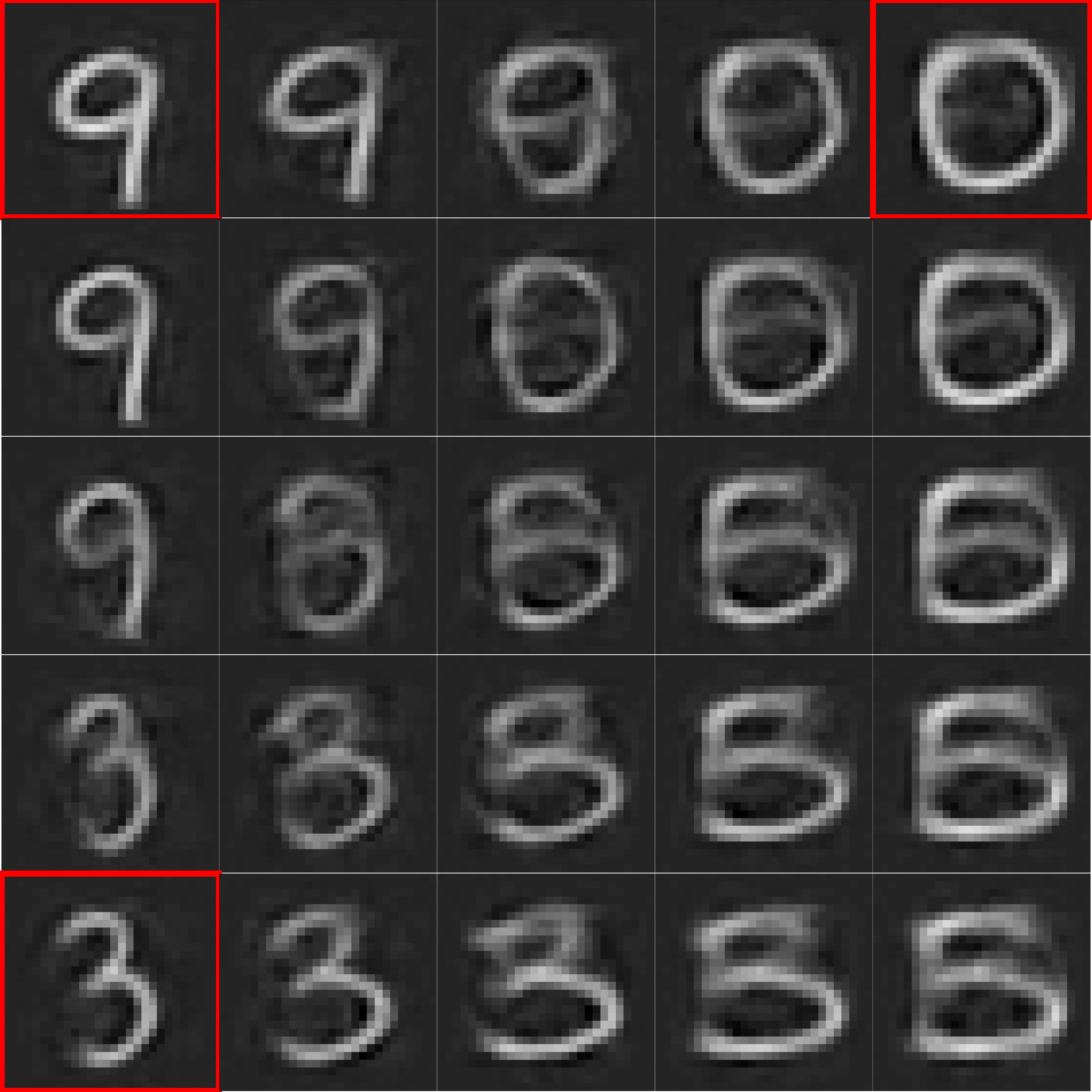}  
		\caption{\texttt{MNIST} - VAE}
		\label{fig:vaemnistinterpolations8}
	\end{subfigure}
	\begin{subfigure}{.32\textwidth}
		\centering
		\includegraphics[width=\linewidth]{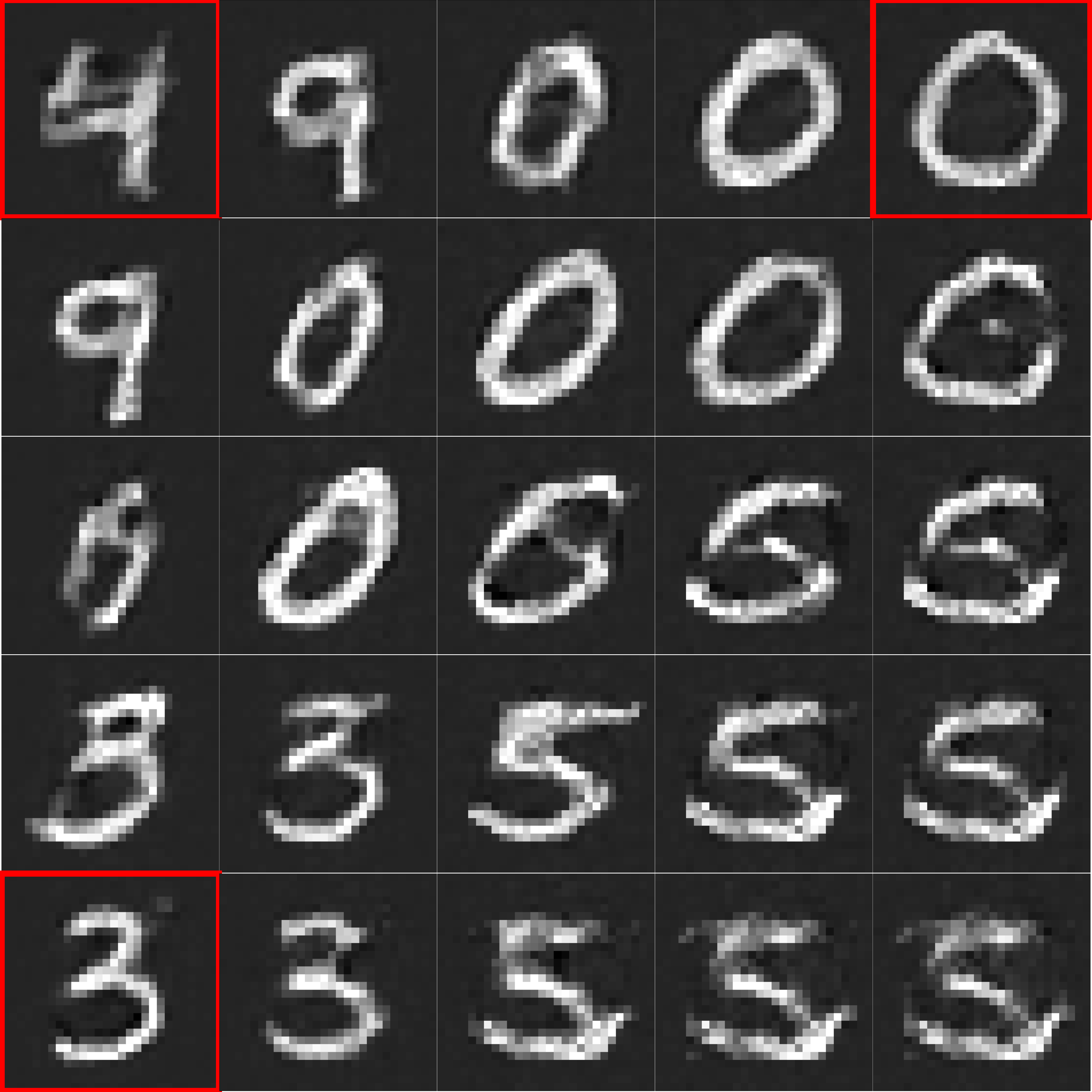}  
		\caption{\texttt{MNIST} - GAN}
		\label{fig:ganmnistinterpolations8}
	\end{subfigure}
	\begin{subfigure}{.32\textwidth}
		\centering
		\includegraphics[width=\linewidth]{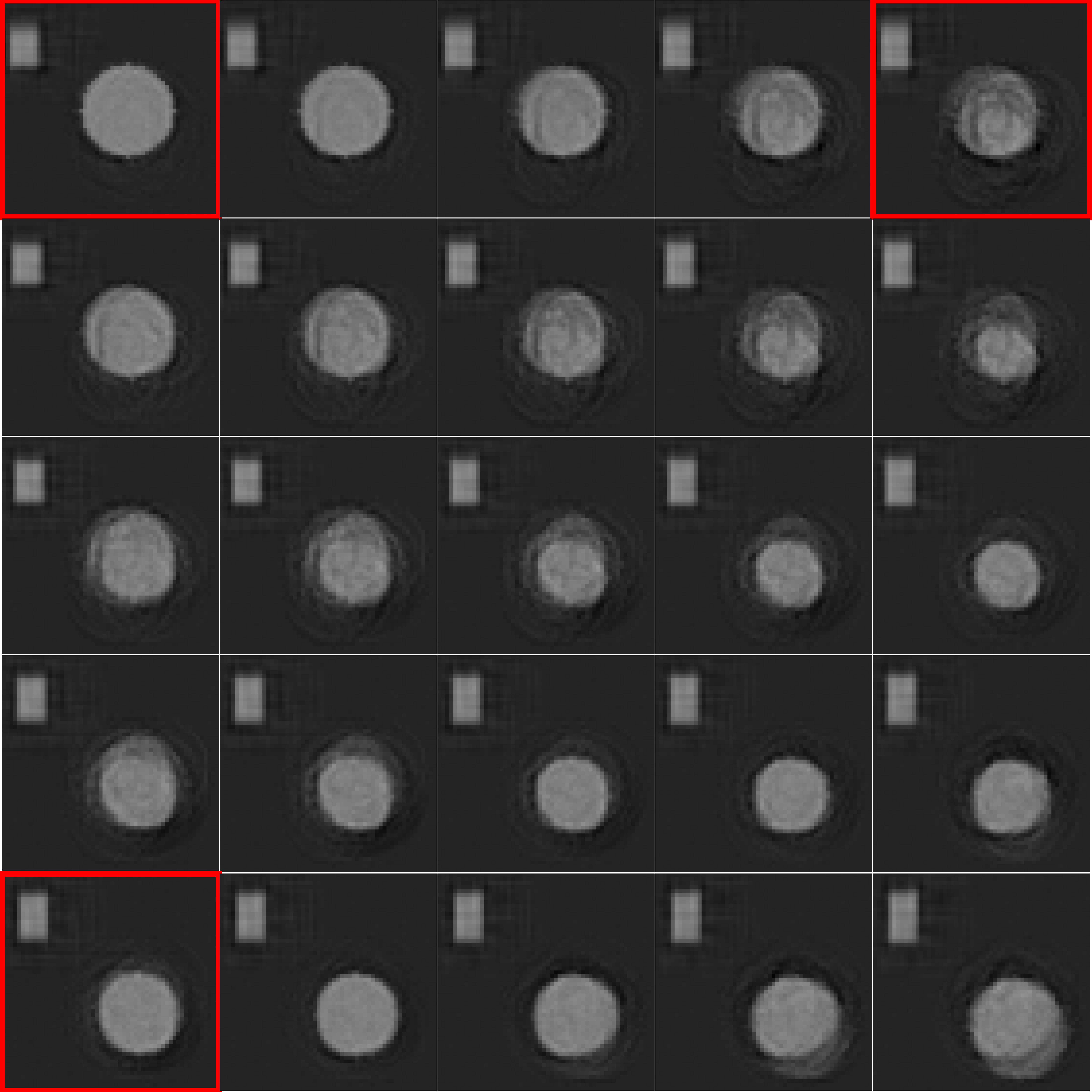}  
		\caption{\texttt{Shapes} - AE}
		\label{fig:aetrainnoneinterpolations10}
	\end{subfigure}
	\begin{subfigure}{.32\textwidth}
		\centering
		\includegraphics[width=\linewidth]{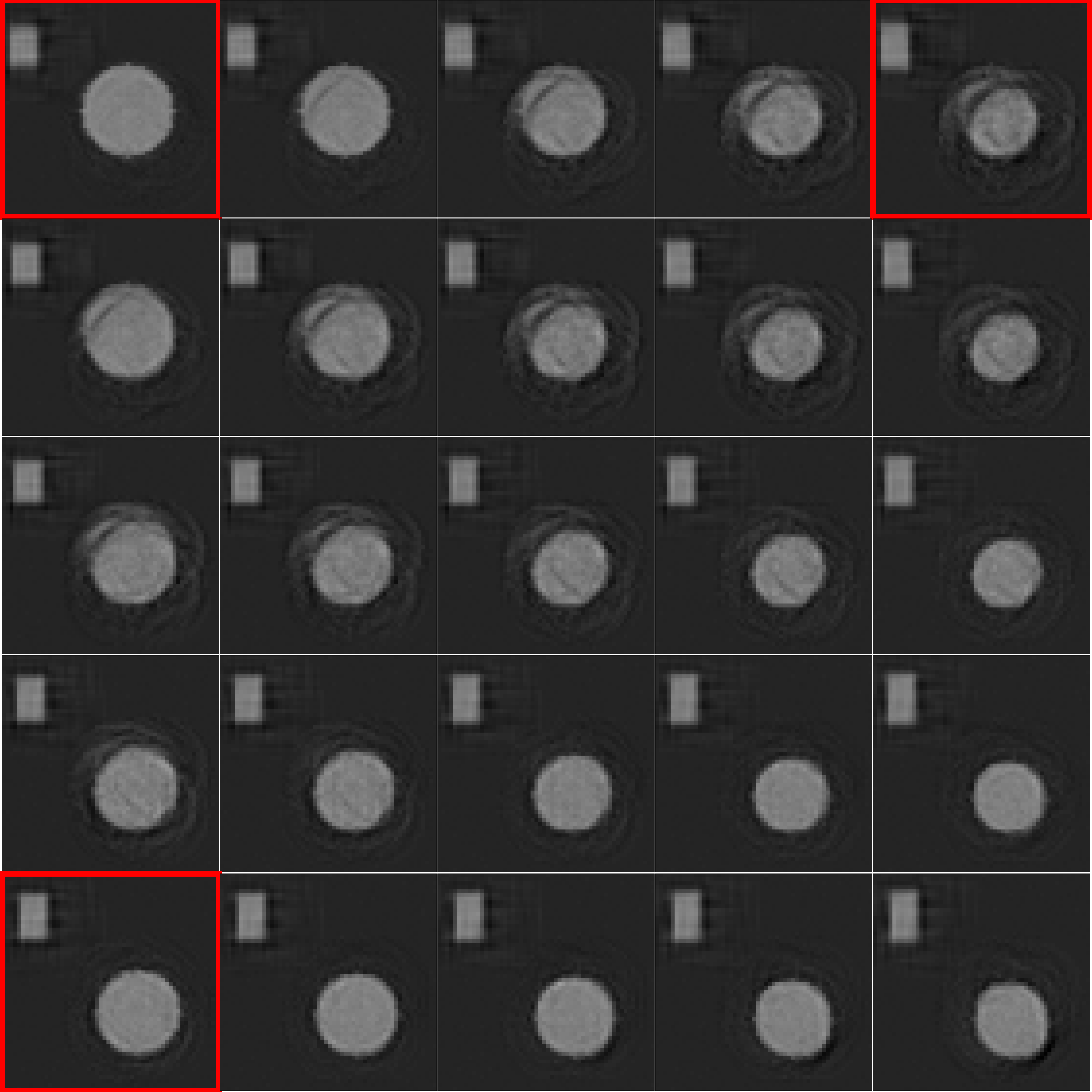}  
		\caption{\texttt{Shapes}  - VAE}
		\label{fig:vaetrainnoneinterpolations10}
	\end{subfigure}
	\begin{subfigure}{.32\textwidth}
		\centering
		\includegraphics[width=\linewidth]{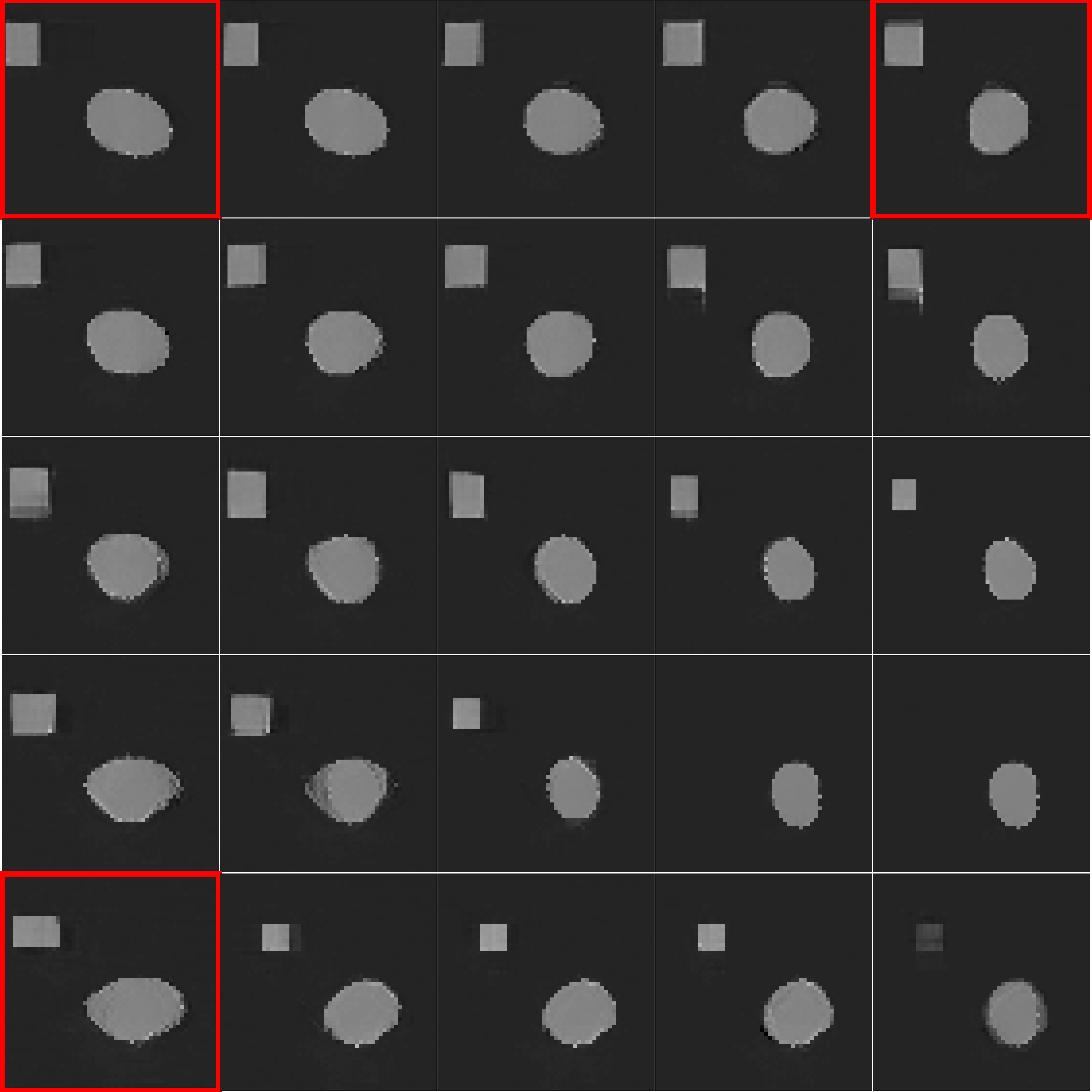}  
		\caption{\texttt{Shapes}  - GAN}
		\label{fig:gantrainnoneinterpolations10}
	\end{subfigure}

	\centering
	\caption{Interpolation ability of an AE, VAE and GAN.  The highlighted top left, bottom left and top right latent space values  were chosen  close to  the test dataset  and the other images are computed via linear combinations in the latent space.}
	\label{fig:interpolations}
\end{figure}

 \changes{We consider interpolating between points in the latent space, testing property B1. We hope to see smooth transitions between interpolated images, and that generated images are similar to those seen in training.  We take three images from the test data, $x_1, x_2$ and $x_3$, find  $z_1, z_2$ and $z_3$, their encodings in the latent space, using \eref{encode} and then plot interpolations  $ G( z_1 + \alpha_1(z_2-z_1)+\alpha_2(z_3-z_1))$ for  $\alpha_1, \alpha_2\in[0,1]$.  Figure  \ref{fig:interpolations} shows one example for each model and dataset for  $\alpha_1, \alpha_2=\{0,0.25,0.5,0.75,1\}$. In the AE and VAE, you see transitions that are smooth but blurry. The GAN images appear sharper but some outputs are not similar to training data examples, for example, in Figure  \ref{fig:gantrainnoneinterpolations10} there are a set of images that contain no rectangle.} These images could be evidence of a discriminator failure:  the discriminator has not yet learnt that these images are not similar to the training set.

	\subsubsection{Discussion}
As expected, none of the three generator models, AE, VAE and GAN, fulfil Property A and B fully. For A,  the GAN does poorly in the reconstruction results of Figures \ref{fig:violinplots} and \ref{fig:mnist_shapes_reconstructions}. The lack of encoder makes this  more challenging. There is evidence of mode collapse, where parts of the training data are not well reconstructed and discriminator failure, where the images produced are not realistic, see Figures \ref{fig:mnist_shapes_reconstructions} and \ref{fig:interpolations}. The VAE does consistently better, demonstrated by the lower EMD between generated and test data in Figure  \ref{fig:distributionscompare}. The lack of prior on the AE, and thus a known area of the latent space to sample from, is a problem. Figure~\ref{fig:farfromdistribution} demonstrates that sampling from the wrong area of the latent space gives poor results. 

Pulling apart the cause of a failure to recover an image is  difficult. It could be that the image is not in the range of the generator, a failure of Property A, or that  the image is in the range of the generator but the image cannot  be recovered using descent methods, a failure of Property B. For property B1, the mathematical properties of continuity or differentiability of a network, depend on the architecture. The  interpolations in Figure  \ref{fig:interpolations} show some evidence of large jumps between images in the GAN cases, but in general the interpolations are reasonable. For both the GAN and the VAE, in Figure~\ref{fig:encodingprojectionplots}, the encodings of the test images in the latent space seem to match the prior, Property B2. 

\section{Numerical Results for Inverse Problems \label{inv_prob_numerics}}

In this section, we apply AE, VAE and GAN models, evaluated in the previous Section, on  three inverse problems. Firstly \textit{tomography},  the X-ray transform~\cite{Natterer2001} with a parallel beam geometry. Secondly, \textit{deconvolution} with a $5\times5$ Gaussian kernel. Lastly, \textit{compressed sensing} where $y=Ax$ is an under-determined linear system  where $A$ is an   $\mathbb{R}^{m\times n} $ Gaussian random matrix, $x \in \mathbb{R}^{n}$ is a vectorised image and  $n\gg m$,  see for example~\cite{Davenport2012}.	 In each case  zero-mean  Gaussian noise  with standard deviation $\sigma$ is added to the data. The forward operators were implemented using the operator discretisation library (ODL)~\cite{Adler2017} in Python, accessing scikit--learn~\cite{scikit-learn} for the tomography back-end.

\changes{We} consider  variational regularisation methods in the form of \eref{objective} \changes{and \eref{gen_regulariser} with $L_y(Ax)= \|Ax-y\|_2^2$}. \changes{To match the literature themes, we compare three different methods: \textsf{hard}, $F(u)=\iota_{\{0\}}(u)$ and $R_\mathcal{Z}(z)=\|z\|_2^2$; \textsf{relaxed}, $F(u)=\|u\|_2^2$ and $R_\mathcal{Z}(z)=\mu\|z\|_2^2$;  and  \textsf{sparse}, $F(u)=\|u\|_1$ and $R_\mathcal{Z}(z)=\mu\|z\|_2^2$}, where $\mu$ is an additional regularisation parameter. We compare with regularisers independent of the generator:  \textsf{Tikhonov} regularisation,
$R_G(x)=\|x\|_2^2$, for the convolution and tomography examples  and TV regularisation~\cite{Rudin1992}, for the compressed sensing example.

The optimisation algorithms are given in the appendix.    \textsf{Hard}  and \textsf{Tikhonov}   are optimised using gradient descent with backtracking line search, Algorithm \ref{GD}. TV regularisation is implemented using the Primal-Dual Hybrid Gradient method~\cite{Chambolle2011}.
For \textsf{relaxed}, alternating gradient descent with backtracking is used, see Algorithm \ref{AGD}. \changes{Finally, for \textsf{sparse}, the 1-norm is not smooth, and so   Proximal Alternating Linearised Minimisation (PALM)~\cite{Bolte2014} with backtracking, Algorithm \ref{PALM}, is used  to optimise the equivalent formulation   $ \min_{u\in\mathcal{X}, z \in\mathcal{Z}}\|A(G(z)+u)-y\|_2^2+\lambda\left(\|u\|_1+\mu\|z\|_2^2\right)$.
In all cases, initial values  are  chosen from a standard normal distribution.}

\subsection{Deconvolution} 

 \changes{Figure   \ref{fig:ganmnistone_initialisation} shows solutions to the deconvolution inverse problem  with added Gaussian noise (standard deviation $\sigma=0.1$) on  the \texttt{MNIST} dataset. We test the \textsf{relaxed}, \textsf{hard} and \textsf{sparse} methods    with  a GAN against \textsf{Tikhonov} and try a range of regularisation parameters. Each reconstruction used the same realisation of noise affecting the data.  \textsf{Hard}  gives good PSNR results despite not  reaching the Morozov discrepancy value, the expected value of the L2 norm of the added Gaussian noise~\cite{Tikhonov1995}.  For the \textsf{hard} constraints reconstructions are restricted to the range of the generator and we do not expect the data discrepancy to go to zero as $\lambda$  decreases.   In the \textsf{relaxed} and \textsf{sparse} constrained reconstructions, for  smaller values of $\lambda$ the solutions tend towards a least squares solution which fits  the noise and is affected by the ill-posedness of the inverse problem.  The additional variation in the choice of $\mu$, as shown by the additional coloured dots, has little effect for  smaller values of $\lambda$. }

\changes{Figure~\ref{fig:vaemnistone}  again shows a deconvolution problem with added Gaussian noise (standard deviation $\sigma=0.1$)  on the \texttt{MNIST} dataset. We choose the  \textsf{hard}  reconstruction for the three different generator models and show three random initialisations in $\mathcal{Z}$. Regularisation parameters were chosen to maximise PSNR.  The best  results are given by the AE and the VAE. The GAN has failed to find a good value in the latent space to reconstruct the number three. The  choice of initial value of $z$  significantly affects the outcome of the reconstruction in the GAN case.    }
\changes{
\begin{figure}
	\centering

	\begin{subfigure}{.47\linewidth}
		\centering
		\includegraphics[width=\linewidth]{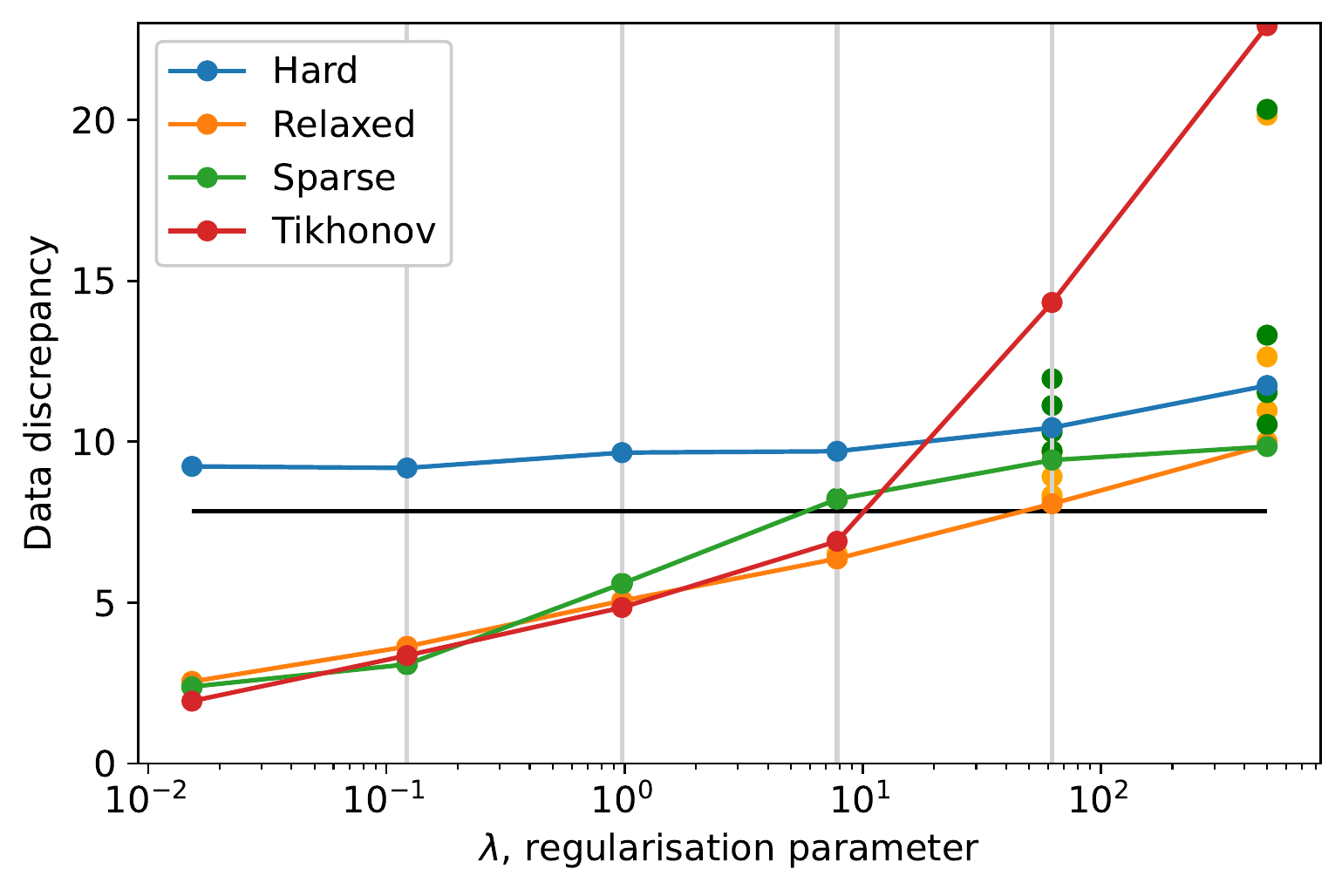}  
		\label{fig:ganxtikone_initialisation}
	\end{subfigure}
	\begin{subfigure}{.47\linewidth}
		\centering
		\includegraphics[width=\linewidth]{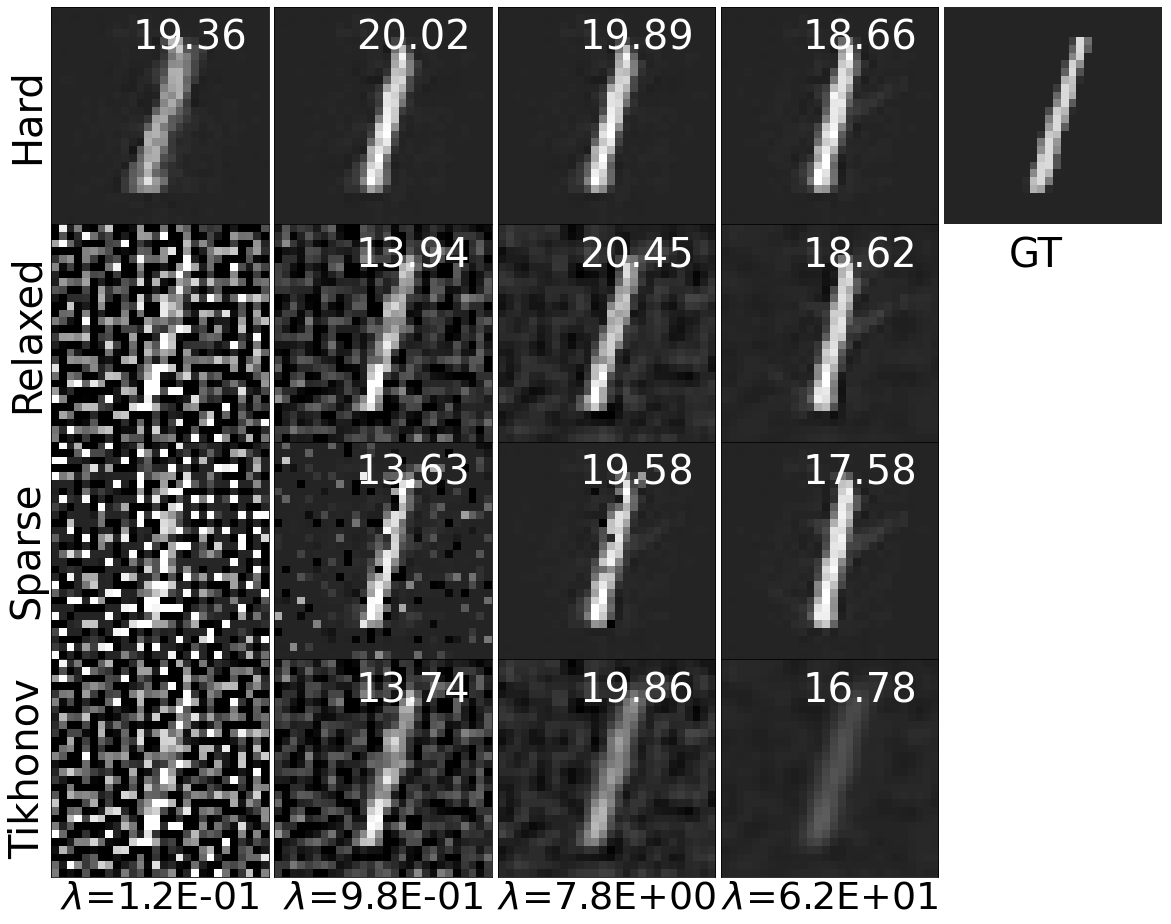}  
		\label{fig:ganzoptone_initialisation}
	\end{subfigure}

	\centering
	\caption{\changes{Solution of the deconvolution problem on \texttt{MNIST} with an eight-dimensional GAN. The plot shows the L2 reconstruction loss against regularisation parameter choice $\lambda$ in comparison with the Morozov discrepancy value in black. Differing choices for $\mu$ are plotted as additional markers.    The image plots correspond to the parameter values shown by the grey lines and include the PSNR values.}}
	\label{fig:ganmnistone_initialisation}
\end{figure}

\begin{figure}
	\centering
		\begin{subfigure}{.1\linewidth}
		\centering
		\includegraphics[width=\linewidth]{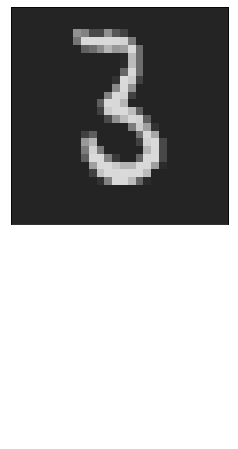}  
		\caption{GT}
		\label{fig:vaezoptone1}
	\end{subfigure}
	\begin{subfigure}{.29\linewidth}
		\centering
		\includegraphics[width=\linewidth]{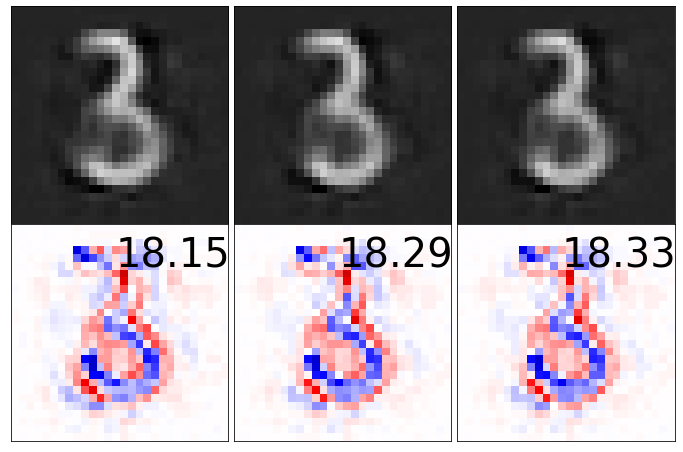}  
		\caption{AE}
		\label{fig:vaezoptone2}
	\end{subfigure}
	\begin{subfigure}{.29\linewidth}
		\centering
		\includegraphics[width=\linewidth]{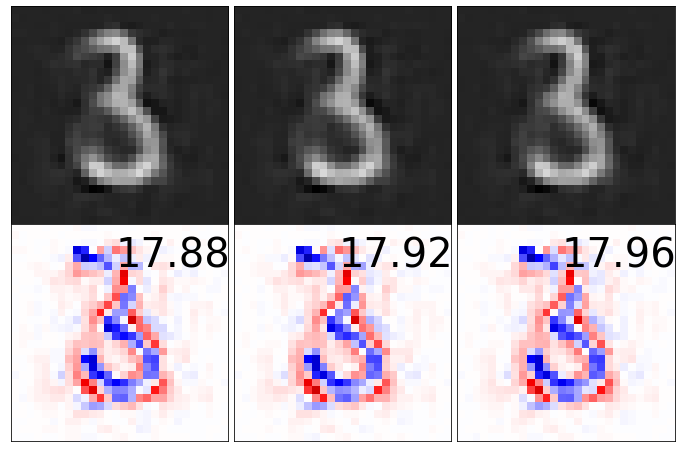}  
		\caption{VAE}
		\label{fig:vaexsoftone3}
	\end{subfigure}
	\begin{subfigure}{.29\linewidth}
		\centering
		\includegraphics[width=\linewidth]{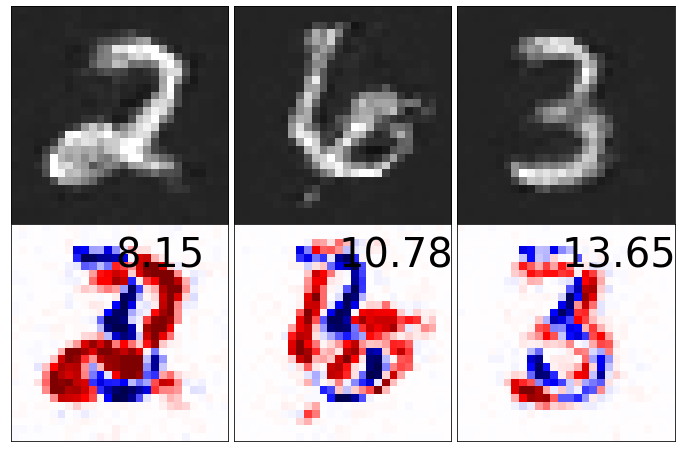}  
		\caption{GAN}
		\label{fig:vaezsparseone4}
	\end{subfigure}

	\centering
	\caption{\changes{Comparisons between the three generators, with eight-dimensional latent space, for the deconvolution problem. Reconstructions use the \textsf{hard} method. The plot shows  3 different initialisations for each generator. The ground truth (GT) is given on the left, the top line shows the reconstruction and the bottom line the residuals with the PSNR values.}}
	\label{fig:vaemnistone}
\end{figure}}

\subsection{Compressed sensing}\changes{
Consider the compressed sensing inverse problem ($m=150$ measurements) with added Gaussian noise (standard deviation $\sigma=0.05$)  on \texttt{MNIST} images.  We choose  regularisation parameters that optimise PSNR over 20 test images.  Figure~\ref{MNIST-psnr-fixed} includes a table with the PSNR results on an additional 100 test images.  Due to the cartoon like nature of the \texttt{MNIST} digits, TV regularisation is particularly suitable, however VAE and AE \textsf{hard} and VAE \textsf{relaxed} are competitive with TV.   For more context, example plots  for the the VAE  and  TV reconstructions are given in Figure~\ref{MNIST-psnr-fixed}. }

\changes{	To give an indication of computational cost, \textsf{Tikhonov} reconstruction on the  compressed sensing  inverse problem on the \texttt{MNIST} dataset took on average 32 iterations of backtracking until the relative difference between iterates was less than $10^{-8}$. In comparison, the \textsf{hard}  and \textsf{relaxed}  took on average 54  and 325 backtracking steps, respectively, without random restarts. The algorithms took up to   1 second for  \textsf{Tikhonov}, 5 seconds for  \textsf{hard}  and 10 seconds for  \textsf{relaxed}.

\begin{figure}
	\centering
\begin{tabular}{c|cccc}
	\hline 
	& \multicolumn{4}{c}{Generative Model}\\
	Method & AE & VAE & GAN & None \\ 
	\hline 
	
	\textsf{Relaxed}  & 19.31 $ \pm$ 2.26  & 20.07 $ \pm$ 1.66 & 17.12 $\pm$ 1.67 &  \\ 
	
	\textsf{Sparse} & 19.52 $ \pm$ 2.72 & 21.08 $ \pm$ 3.16 & 17.06 $ \pm$ 2.5& \\ 

	\textsf{Hard}  & 21.84 $ \pm$ 4.09 & \textbf{22.33 $\pm$ 4.17} & 16.93 $\pm$ 2.57& \\ 

	TV&  &  &   & 21.54 $ \pm$ 1.32\\ 
	\hline 
\end{tabular} 
	\includegraphics[width=\linewidth]{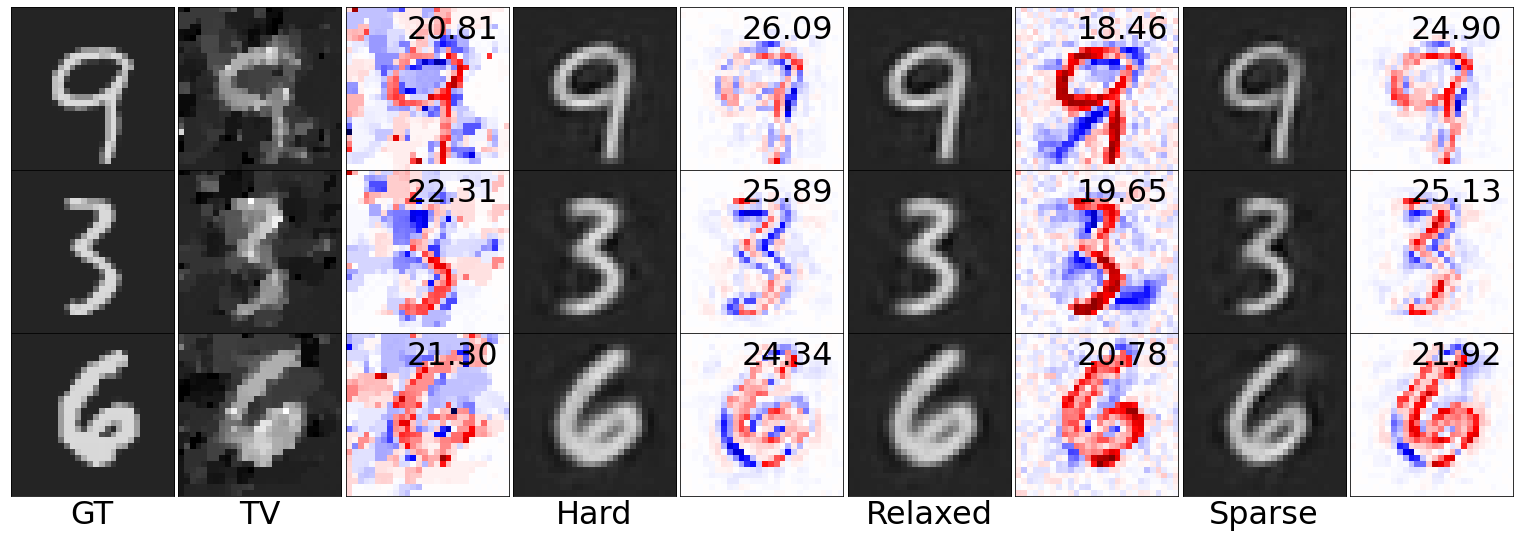}
    \caption{\changes{Results compare the three different regularisers and three different methods against the unlearned TV reconstruction on the compressed sensing inverse problem. The table shows the mean and standard deviations of PSNR values of 100 reconstructions.  The plots show three example solutions, comparing the VAE reconstructions to TV reconstruction. The left column shows the ground truth, even columns the reconstructed images and odd columns the residuals with the PSNR values.}}
    \captionsetup{labelformat=andtable}
\label{MNIST-psnr-fixed}
\end{figure}}
\subsection{Tomography} \changes{
Taking the tomography inverse problem with added Gaussian noise (standard deviation $\sigma=0.1$),  Figure  \ref{SHAPES-psnr-fixed} includes a table which gives the average and standard deviation for the PSNR of 100 reconstructed \texttt{Shapes} images. The regularisation parameters were set to maximise the PSNR  over a separate dataset of 20 test images. The GAN has a particularly poor performance but the AE and VAE results are all competitive with TV.   Example reconstructions for the AE methods and TV reconstruction are given in Figure~\ref{SHAPES-psnr-fixed}. The generative regulariser gives a clear rectangle and circle while the TV reconstruction gives shapes with unclear outlines and blob like artefacts. } \changes{In terms of computational cost, \textsf{Tikhonov}  took on average 157 iterations of backtracking until the relative difference between iterates was less than $10^{-8}$. In comparison, the \textsf{hard}  and \textsf{relaxed}  took on average 37 and 255 iterations, respectively, without random restarts. The algorithms took up to  40 seconds for  \textsf{Tikhonov}, up to  12 seconds for \textsf{hard} and up to 60 seconds for  \textsf{relaxed}.  }
\changes{\begin{figure}
	\centering
	\begin{tabular}{c|cccc}
		\hline 
		& \multicolumn{4}{c}{Generative Model}\\
		Method & AE & VAE & GAN & None \\ 
		\hline 
		
		\textsf{Telaxed}  & 30.70 $\pm$ 1.59 & 28.79 $\pm$ 0.71 & 24.20 $\pm$ 0.49 &  \\ 
		
		\textsf{Sparse} & 33.12 $\pm$ 0.80 & 33.09 $\pm$ 0.89 & 22.72 $\pm$ 1.80& \\ 
		
		\textsf{Hard}  & \textbf{33.17 $\pm$ 0.80} & 32.92 $\pm$ 0.95  & 22.92 $\pm$ 2.10& \\ 
		
		TV&  &  &   & $29.94 \pm 0.75$ \\ 
		\hline 
	\end{tabular} 
	\includegraphics[width=\linewidth]{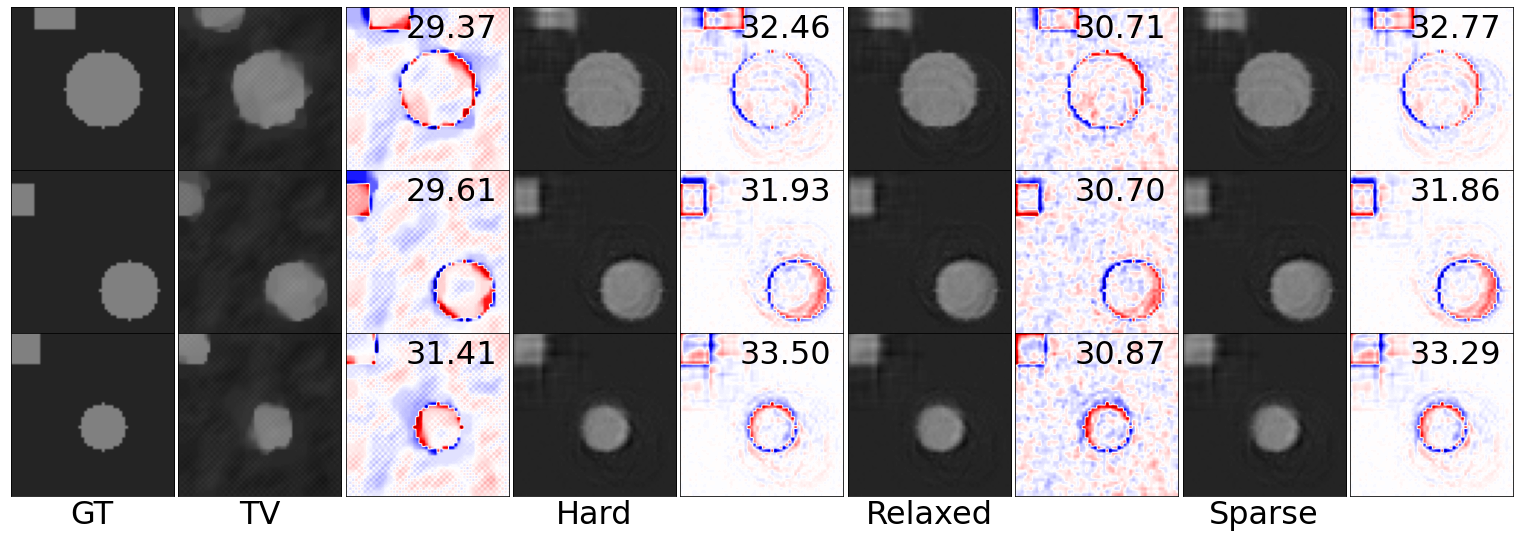}

    \caption{\changes{Results compare the three different regularisers and three different methods against the unlearned TV reconstruction on the tomography inverse problem. The table shows the mean and standard deviations of PSNR values of 100 reconstructions.  The plots show five example reconstructions, comparing the AE reconstructions to TV reconstruction. The left column shows the ground truth, even columns the reconstructions and odd columns the residuals with the PSNR values.}}
    \captionsetup{labelformat=andtable}
\label{SHAPES-psnr-fixed}
\end{figure}}
\subsection{Out-of-Distribution Testing}
\changes{\begin{figure}
	\centering
		\includegraphics[width=0.6\linewidth]{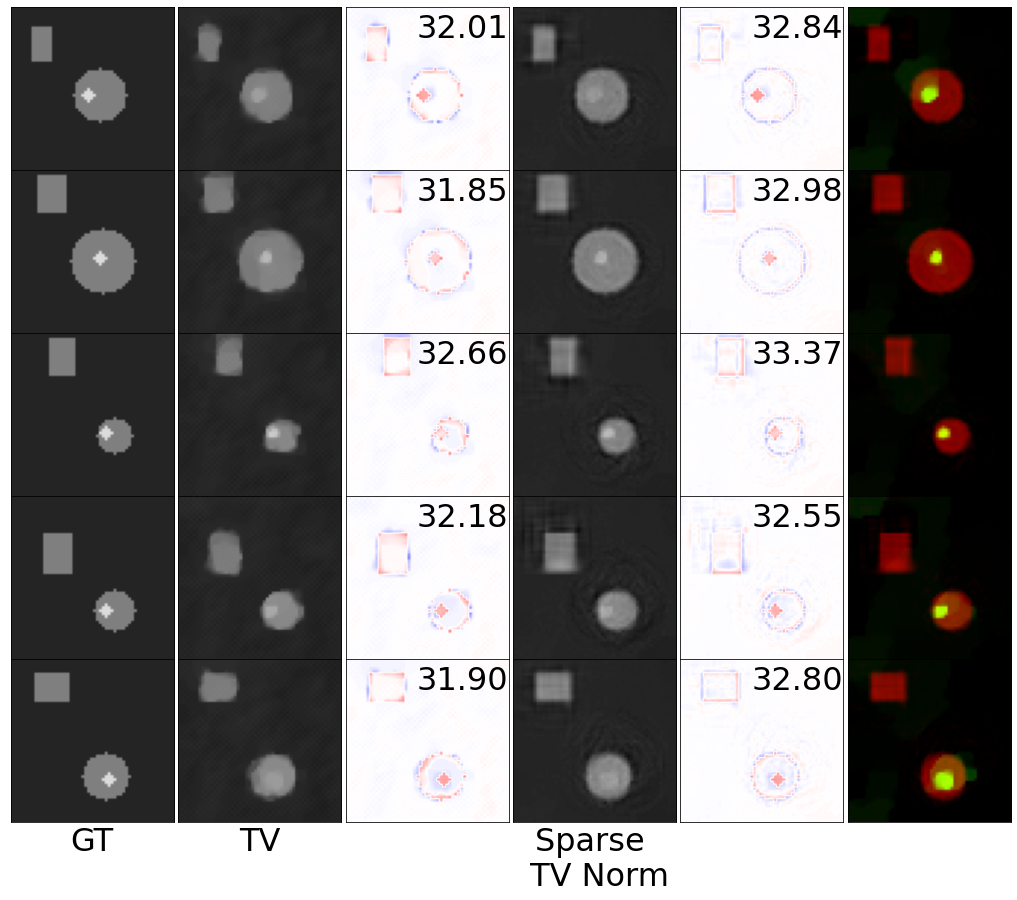}
		\label{fig:bright_tv_sparse}
		\caption{\changes{Tomography inverse problem  on random images from the \texttt{shapes+} dataset. It compares the use of \textsf{sparse}  method, where sparsity is measured in the TV-norm, with a standard TV reconstruction. The generator is a 10 dimensional VAE trained on \texttt{Shapes} images.  In the final column the part of the reconstruction lying in the range of the generator is coloured red and  the sparse addition yellow.}} 
		\label{fig:vaesparsetvpsnrlossfixedbright0}
	\end{figure}}
\changes{ We  augment the \texttt{Shapes} dataset, creating a \texttt{shapes+} dataset, with the addition of a bright spot randomly located in the circle.  We then take the Tomography inverse problem on the \texttt{shapes+} dataset with added Gaussian noise (standard deviation $\sigma=0.05$).  For a generative regulariser we use \textsf{sparse}, with $F(x)=\|\nabla x\|_1$, the TV norm. Crucially, the VAE generator used was trained only on the standard \texttt{Shapes} dataset, without bright spots.     We compare with standard TV reconstruction. The regularisation parameters were chosen to maximise the PSNR on 20 ground truth and reconstructed images. The mean PSNR over 100 test images for the \textsf{sparse} case is 32.83 with standard deviation  0.65 and for the TV reconstruction is  32.01 with standard deviation  0.67.  Figure~\ref{fig:vaesparsetvpsnrlossfixedbright0} show five reconstructions.    The \textsf{sparse} deviations allow reconstruction of the bright spot demonstrating that generative regularisers can also be effective on images close to, but not in, the training distribution.}

\subsection{\changes{\texttt{FastMRI} Dataset}} \changes{
 
We also train a VAE to produce knee \texttt{FastMRI} images. The VAE architecture is  based on  Narnhofer \etal. The \texttt{FastMRI} knee dataset contains data 796 fully sampled knee MRI magnitude images~\cite{Knoll2020a, Zbontar2018}, without fat suppression.  We extract 3,872 training and 800 test ground truth  images from the dataset, selecting images from near the centre of the knee,  resize the images to $128\times128$ pixels  and  rescale to the pixel range $[0,1]$.   The \texttt{FastMRI} VAE models took approximately 12 hours to train \changes{on the same system as above}.}

\changes{
Results for the tomography inverse problem with added Gaussian noise of varying standard deviation are given in Figure	\ref{fig:knees_tomo}. For each image and noise level, the same noise instance is used for each reconstruction method, and additionally, for each method, a range of regularisation parameters are tested, and the reconstruction with the best PSNR value chosen. The plot shows how the PSNR values, averaged over 50 test images, vary with noise level.  The \textsf{relaxed} method gives the best PSNR values, outperforming \textsf{Tikhonov} although with larger variance.  The \textsf{sparse} method curve has a similar shape to Tikhonov, but performs consistently worse suggesting that this choice of deviations from the generator is not suited to this dataset, generator or inverse problem.   We see that for the \textsf{hard} method the results are consistent across the range of noise levels, not improving with reduced noise.  The example images reflect the data with the \textsf{hard} reconstruction doing comparatively better with larger noise levels, but the \textsf{relaxed} method capturing more of the fine details at the lower noise level. }

\changes{

\begin{figure}
	\centering
	\begin{subfigure}{.55\linewidth}
		\centering
		\includegraphics[width=\linewidth]{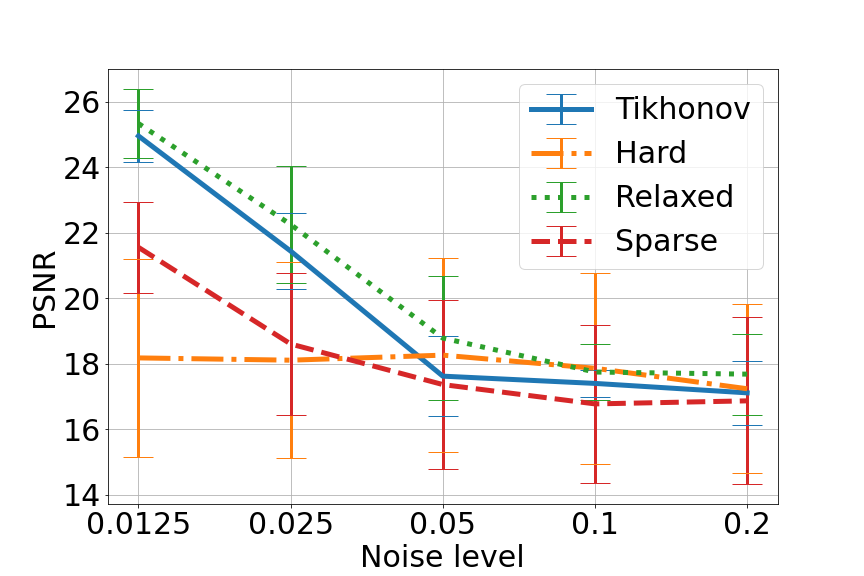}  
		\label{fig:knees_z_opt_237}
	\end{subfigure}	
	\begin{subfigure}{.43\linewidth}
		\centering
		\includegraphics[width=\linewidth]{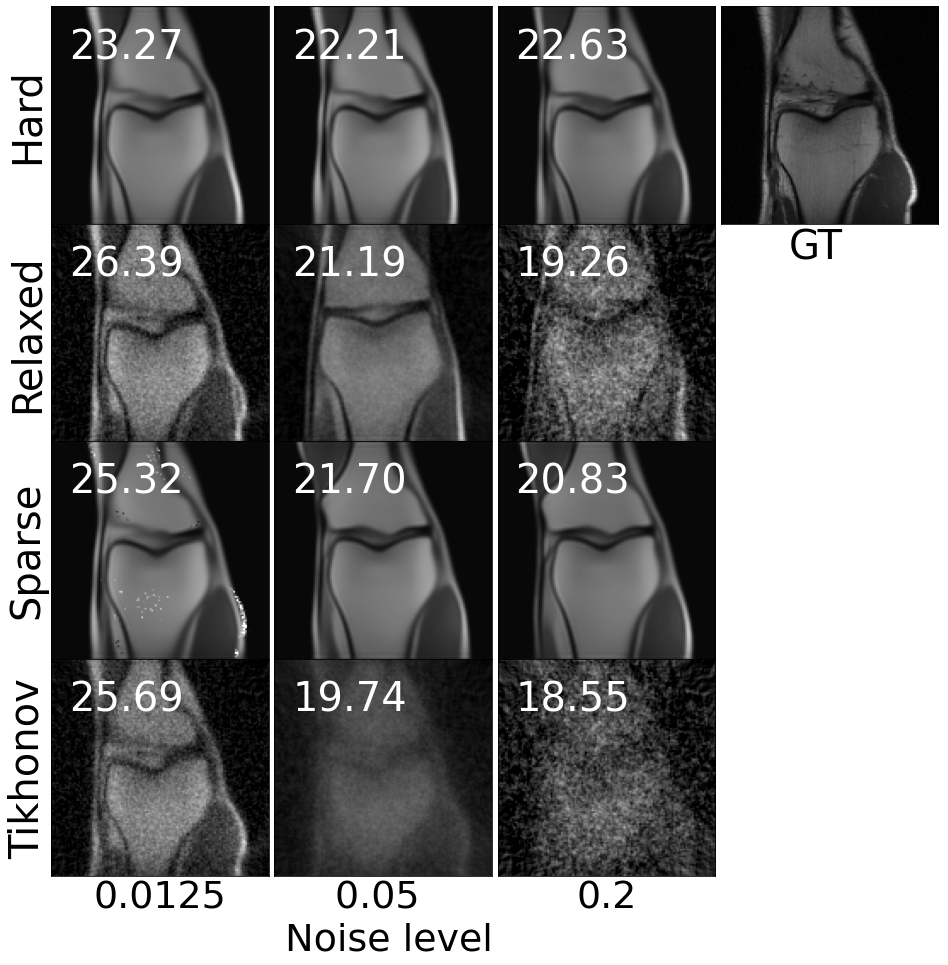}  
		\label{fig:knees_tik_237}
	\end{subfigure}
	\centering
	\caption{\changes{Reconstructions of the tomography inverse problem with additive Gaussian noise of varying standard deviations. Regularisation parameters are chosen to maximise PSNR value for each image.   The left shows the average PSNR values, taken over 50 test images, with the standard deviation in the error bars and the right shows one particular example reconstruction with PSNR values.}}
	\label{fig:knees_tomo}
\end{figure}

}
	\section{Summary, Conclusions and Future Work  \label{conclusions}}

This paper looked at the use of a generator, from a generative machine learning model, as part of the regulariser of an inverse problem. We named these \textit{generative regularisers}. Generative regularisers link the theoretically well-understood methods of variational regularisation with state-of-the-art machine learning models. The trained generator outputs data similar to  training data and the regulariser restricts solutions (close) to the range of the generator. The cost of these generative regularisers is in the  need for generative model training, the requirement for a large amount of training data and \changes{the difficulty of} the resulting non-convex optimisation scheme. Weighing up the costs and benefits will depend on the inverse problem and the availability of data.  

We compared three different types of generative regularisers which either restrict solutions to exactly the range of the generator or allow small or sparse deviations. We found that in simpler datasets \changes{the restriction to  the} range of the generator was successful. Where the ground truth was \changes{more complex} then allowing small deviations produced the best results.  A key benefit of generative regularisers over other deep learning approaches is that paired training data is not required, making the method flexible to changes in the forward problem. We demonstrated the use of generative regularisers on  deconvolution and fully sampled tomography problems, both with gradually decaying singular values  of the forward operator;  and compressed sensing, with a large kernel and non-unique solutions. 

The training of the generator is crucial to the success of generative regularisers, and a key contribution of this report is a set of desirable properties for a generator.  Numerical tests linked to these properties were  discussed and applied to three generative models: AEs,  VAEs and GANs. None of these models fulfil the criteria completely. We observed known issues such as mode collapse and discriminator failure in the GAN, blurry images in the VAE and the lack of a prior in the AE. In the inverse problem experiments in this paper the AE and the VAE yielded the most consistent results. The success of the AE, despite the lack of prior on the latent space, surprised us. We suspect the implicit regularisation on the model from the architecture and initialisations helped making the AE a usable generator. The GAN models did worst in the inverse problem examples: they generally seemed more sensitive to initialisation  of the non convex optimisation, making the optimal point in the latent space  difficult to recover.

This paper focused on training the generative model first and then subsequently using the generative regularisation to solve inverse problems. The benefit of this split approach is that the model does not need retraining if there are changes in the forward problem. A further advantage is that the field of generative modelling is growing quickly and any improvements to generators can be directly plugged into these methods. An interesting direction for future work would be to consider training (or refining) generative models with a particular inverse problem in mind.

\section*{Acknowledgements}

MD is supported by a scholarship from the EPSRC Centre for Doctoral Training in Statistical Applied Mathematics at Bath (SAMBa), under the project EP/L015684/1. MJE acknowledges support from the EPSRC (EP/S026045/1, EP/T026693/1), the Faraday Institution (EP/T007745/1) and the Leverhulme Trust (ECF-2019-478).
NDFC acknowledges support  from the EPSRC CAMERA Research Centre (EP/M023281/1 and EP/T022523/1) and the Royal Society.

	\printbibliography
 
\appendix
\changes{
\section{Optimisation Algorithms}

\begin{algorithm}[H]
	\caption{Gradient Descent with Backtracking to solve $\min_z f(z)$.}
	\label{GD}
	\begin{algorithmic}[1]
		\State Initialise $z_0$,  $L>0$, $0<\eta_0<1$, $\eta_1>1$. 
		\For{$i = 1,..., K$} 
		\State Let $\tilde{z}(L):=z_{i-1}-\frac{1}{L}\nabla f(z_{i-1})$
		\While{$f(\tilde{z}(L))\geq f(z_{i-1}) -\frac{1}{2L}\|\nabla f(z_{i-1})\|_2^2$}
		\State $L=L\eta_1$	
		\EndWhile
		\State $z_i=\tilde{z}$  and   $L=L\eta_0$.
		\EndFor
	\end{algorithmic}
\end{algorithm}

\begin{algorithm}[H]
	\caption{Alternating gradient descent with backtracking to solve $\min_{z,x} f(z,x)$.}
	\label{AGD}
	\begin{algorithmic}[1]
		\State Initialise $z_0$ and $x_0$, $L_z>0$,$L_x>0$, $0<\eta_0<1$ and  $\eta_1>1$ 
		\For{$i =1,....,K$}
		\State Let $\tilde{z}(L_z):=z_{i}-\frac{1}{L_z}\nabla f(z_{i},x_{i}) $
		\While{$f(\tilde{z}(L_z), x_i)\geq f(z_{i}, x_{i}) - \frac{1}{2 L_z}\|\nabla f(z_{i}, x_{i})\|_2^2$}
		\State $L_z=L_z\eta_1$	
		\EndWhile
		 \State Let $z_{i+1}=\tilde{z}(L_z)$  and then  $L_z=L_z\eta_0$ 
		 \vskip 6pt
		\State Let $\tilde{x}(L_x):=x_{i}-\frac{1}{L_x}\nabla f(z_{i+1}, x_{i})$	
		\While{$f(z_{i+1},\tilde{x}(L_x))\geq f(z_{i+1}, x_{i})-\frac{1}{2L_x}\|\nabla f(z_{i+1}, x_{i})\|_2^2$}
		\State $L_x=L_x\eta_1$
		\EndWhile
		\State Let $x_{i+1}=\tilde{x}(x_L)$ and $L_x=L_x\eta_0$

		\EndFor
	\end{algorithmic}
\end{algorithm}

\begin{algorithm}[H]
	\caption{PALM with backtracking to solve $\min_{z,u} f(z,u)+g_1(z)+g_2(u)$. Define $\mathrm{prox}_{ h}(z)=\arg\min_{x}\{ h(x)+\frac{1}{2}\|x-z\|_2^2\}$. }
	\label{PALM}
	\begin{algorithmic}[1]
		\State Initialise $z_0$,  $u_0$,  $L_z>0$, $L_x>0$,  $0<\eta_0<1$ and $\eta_1>1$.
		\For{$i=1,...K$}

		\State Let $\tilde{z}(L_z):=\mathrm{prox}_{\frac{1}{L_z} g_1}(z_{i}-\frac{1}{L_z}\nabla_z f(z_{i}, u_{i}))$
		\While{$f(\tilde{z}(L_z),u_{i})>f(z_{i},u_{i})+ \nabla_z f(z_{i}, u_{i})^T(\tilde{z}(L_z)-u_{i})+\frac{L_z}{2}\|\tilde{z}(L_z)-z_{i}\|_2^2$}
		\State $L_z=L_z\eta_1$
		\EndWhile
		\State Let $z_{i+1}=\tilde{z}(L_z)$ and then $L_z=L_z\eta_0$
		
		\vskip 6pt
		\State Let $\tilde{u}(L_u):=\mathrm{prox}_{\frac{1}{L_u}  g_2}(u_{i}-\frac{1}{L_u}\nabla_u f(z_{i+1}, u_{i}))$		
		
		\While{$f(z_{i+1},\tilde{u}(L_u))>f(z_{i+1}, u_{i})+ \nabla_u f(z_{i+1}, u_{i})^T(\tilde{u}(L_u)-u_{i})+\frac{L_u}{2}\|\tilde{u}(L_u)-u_{i}\|_2^2$}
		\EndWhile
		\State Let $u_{i+1}=\tilde{u}(L_u)$ and then set $L_u=L_u\eta_0$
		\EndFor
	\end{algorithmic}
\end{algorithm}
}

\section{Generative Model Architectures }
 The architectures for the three different generative models, for the different datasets are given in this Appendix. 
\begin{figure}[h]
	\centering
	\includegraphics[width=0.9\linewidth]{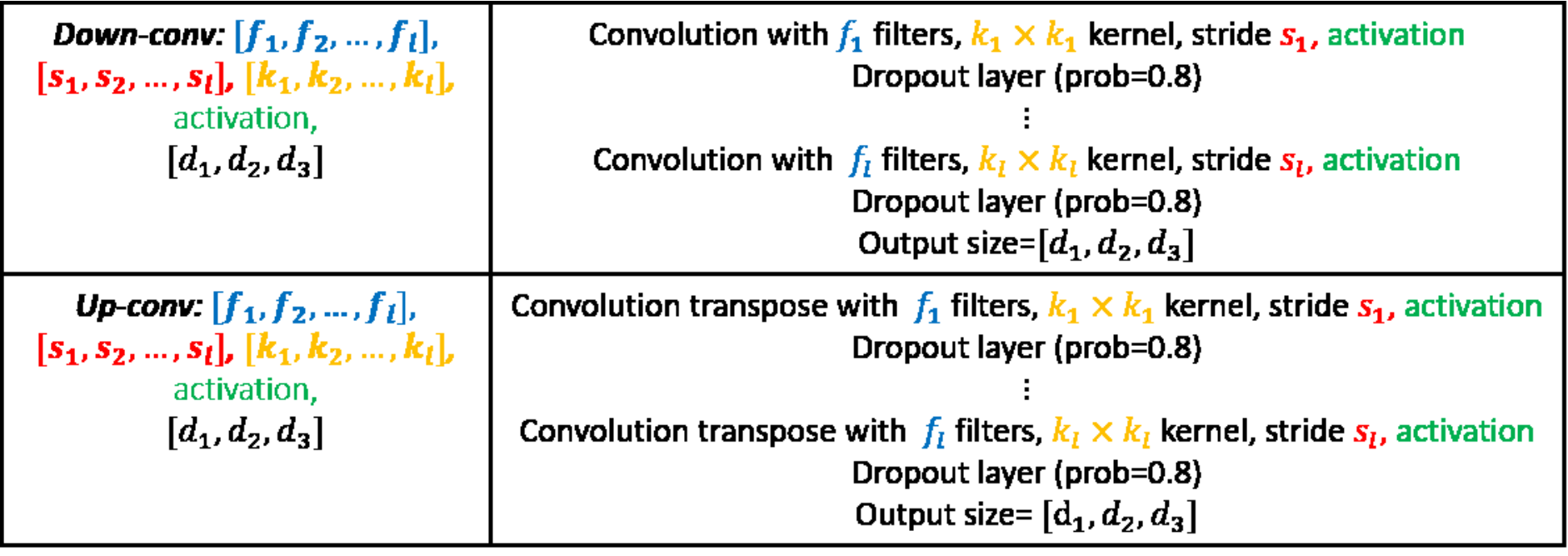}
	\caption{Definitions used in Figure~\ref{fig:mnistarchitectures}, \ref{fig:2dshapesarchitectures} and \ref{fig:kneesarchitectures}.}
	\label{fig:block_architectures}
\end{figure}

\begin{figure}[h]
	\centering
	\includegraphics[width=0.9\linewidth]{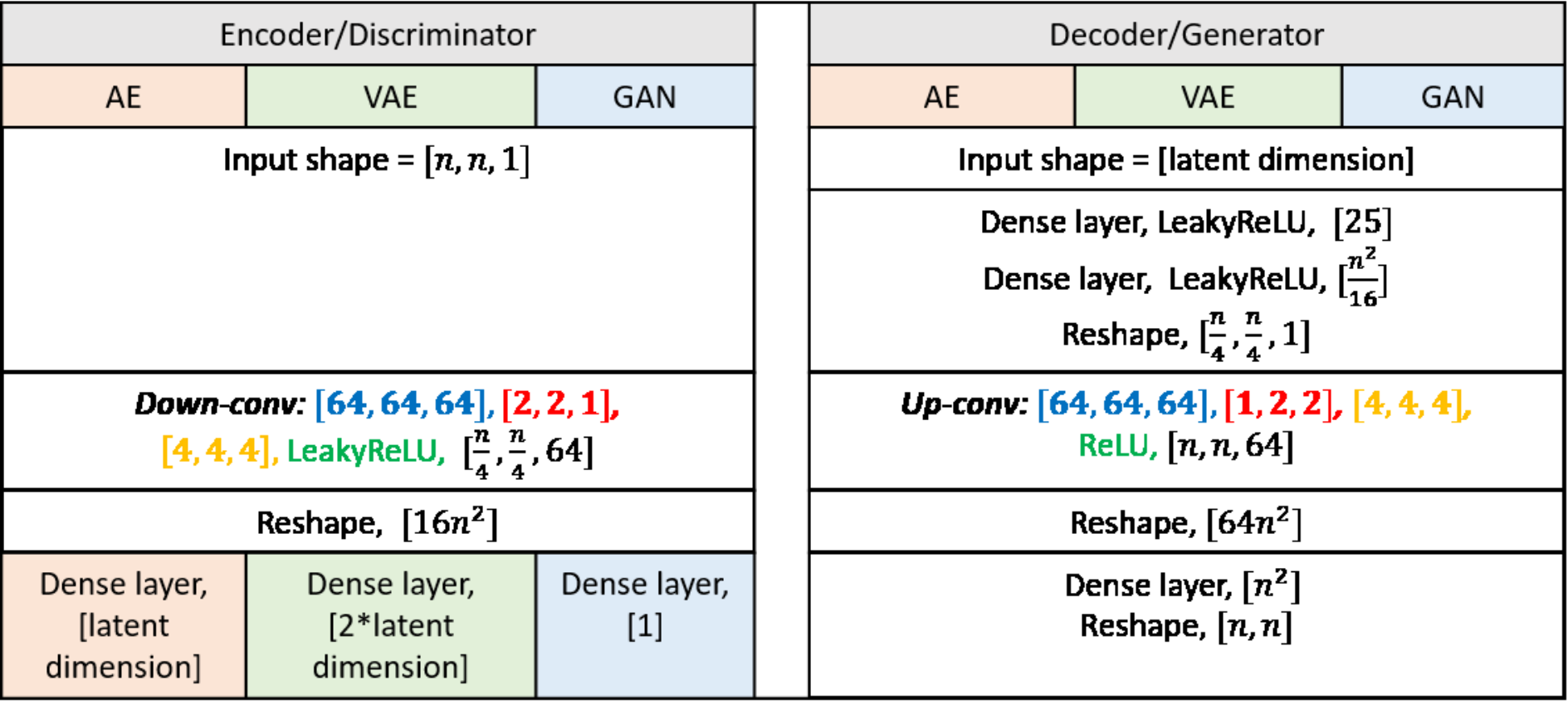}
	\caption{The architectures for the 3 generative models, AE, VAE and GAN, for the \texttt{MNIST} dataset. The convolution block definitions are given in Figure  \ref{fig:block_architectures}.}
	\label{fig:mnistarchitectures}
\end{figure}

\begin{figure}[h]
	\centering
	\includegraphics[width=0.9\linewidth]{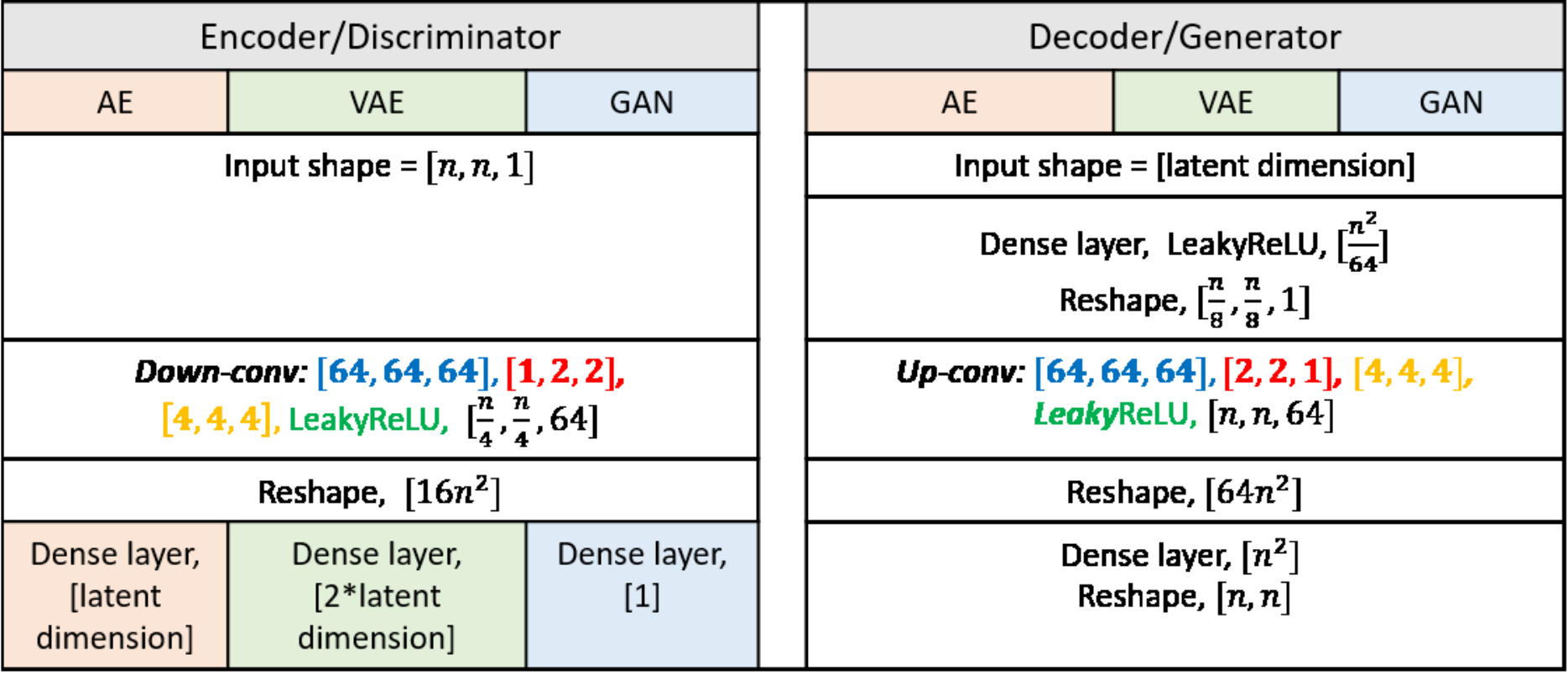}
	\caption{The architectures for the 3 generative models, AE, VAE and GAN, for the \texttt{Shapes} dataset. The convolution block definitions are given in Figure  \ref{fig:block_architectures}.}
	\label{fig:2dshapesarchitectures}
\end{figure}

\begin{figure}[h]
	\centering
	\includegraphics[width=0.9\linewidth]{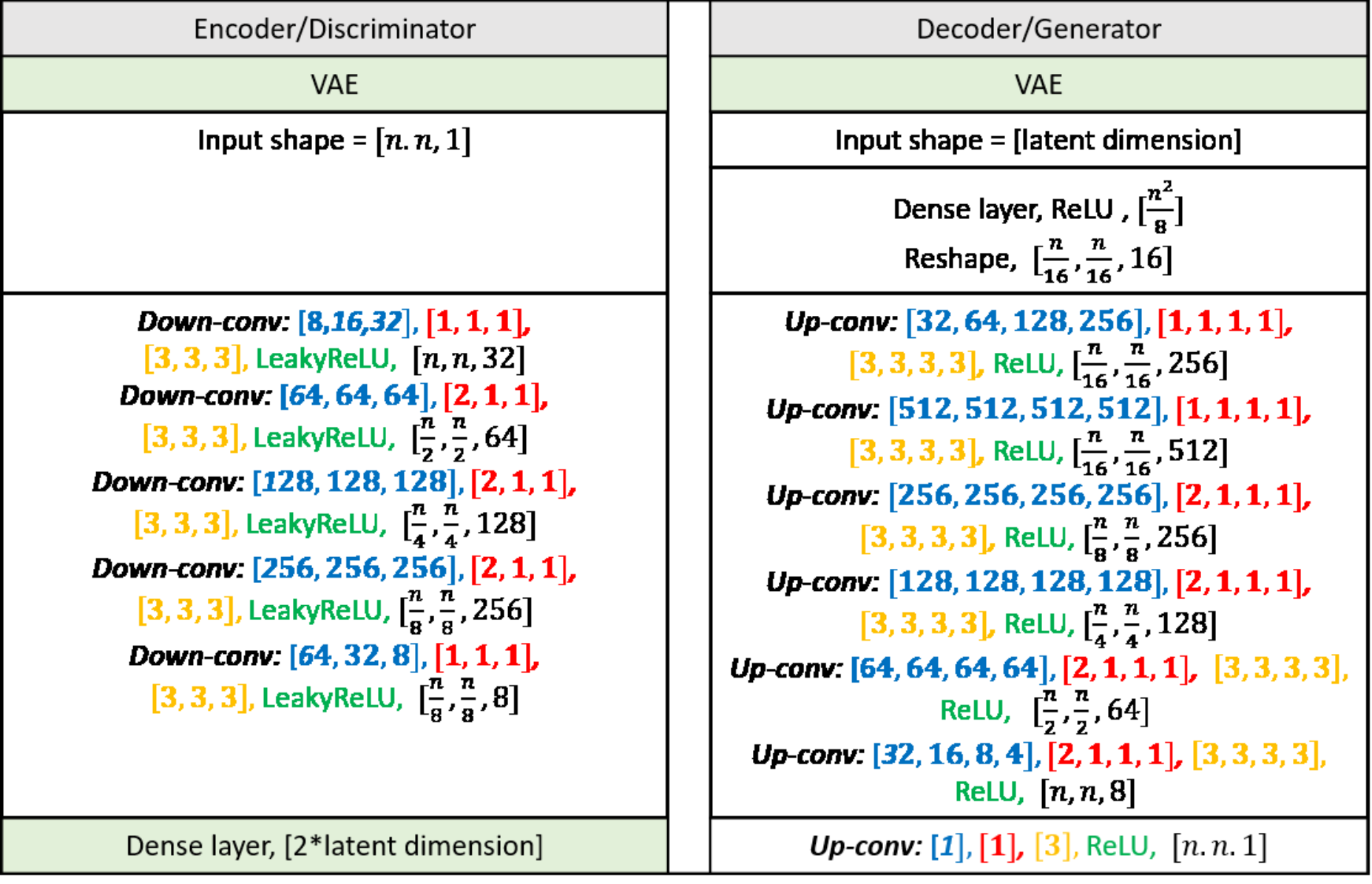}
	\caption{The architectures for the knee dataset VAE. The convolution block definitions are given in Figure  \ref{fig:block_architectures}. }
	\label{fig:kneesarchitectures}
\end{figure}

	\end{document}